\documentclass{aastex63}

\usepackage{savesym}
\savesymbol{tablenum}
\usepackage[alsoload=synchem,alsoload=astro]{siunitx}
\usepackage{siunitx}
\restoresymbol{SIX}{tablenum}
\usepackage[colorinlistoftodos,disable]{todonotes}

\newcommand{\Jansky}{Jy}

\usepackage{amsmath}
\usepackage{natbib}
\usepackage{comment}
\usepackage{multirow}
\usepackage{dblfloatfix}
\usepackage{booktabs} % To thicken table lines
\usepackage{threeparttable}

\hypersetup{linkcolor=blue,anchorcolor=blue,citecolor=blue,filecolor=blue,urlcolor=blue,bookmarksnumbered=true,pdfview=FitB} %

\graphicspath{%
{./HiResFigures/},
{./Figures/}%
}

\shorttitle{IR Extinction at the Galactic Center}
\shortauthors{Stelter and Eikenberry}
\begin{document}

\title{Extinction at the Galactic Center Using Near- and Mid-infrared Broadband Photometry: \\
    A Twist on the Rayleigh-Jeans Color Excess Method}

%: Author Block
\correspondingauthor{Deno Stelter}
\email{deno@ucolick.org}

\author[0000-0003-4549-0210]{R. Deno Stelter}
\affiliation{Center for Adaptive Optics, UC Observatories, UC Santa Cruz, Santa Cruz, CA 95064 USA}
\affiliation{Astronomy Department, UC Santa Cruz, Santa Cruz, CA 95064}
\affiliation{Astronomy Department, University of Florida, Gainesville, FL 32610}

\author[0000-0002-1332-5061]{Stephen S. Eikenberry}
\affiliation{Astronomy Department, University of Florida, Gainesville, FL 32610}
\affiliation{Research Foundation Professor, University of Florida}

%: Submitted, Rec'd block
\received{March 07 2019}
\revised{March 24 2020}
%\accepted{acceptance date}
%\published{published date}
\submitjournal{\apj}

%: Watermark
%\watermark{Last edited 23 March 2020 -- RDS}
\setwatermarkfontsize{0.5in}

\begin{abstract}
We present an extinction map of the inner $\sim$\SI{15}{\arcminute} by {16}{\arcminute} of the Galactic Center (GC) with map `pixels' measuring \SI{5}{\arcsecond} $\times$ \SI{5}{\arcsecond} using integrated light color measurements in the near- and mid-infrared.
We use a variant of the Rayleigh-Jeans Color Excess (RJCE) method first described by~\cite{dustyVeilI} as the basis of our work, although we have approached our problem with a Bayesian mindset and dispensed with point-source photometry in favor of surface photometry, turning the challenge of the extremely crowded field at the GC into an advantage.
Our results show that extinction at the GC is not inconsistent with a single power law coefficient, $\beta=2.03\pm0.06$, and compare our results with those using the Red Clump (RC) point-source photometry method of extinction estimation.
We find that our measurement of $\beta$ and its apparent lack of spatial variation are in agreement with prior studies, despite the bimodal distribution of values in our extinction map at the GC with peaks at \num{5} and \SI{7.5}{mag}. 
This bimodal nature of extinction is likely due to the InfraRed Dark Clouds that obscure portions of the inner GC field.
We present our extinction law and map and de-reddened NIR CMDs and color-color diagram of the GC region using the point-source catalog of IR sources compiled by \cite{dewitt2010}.
The de-reddening is limited by the error in the extinction measurement (typically \SI{0.6}{mag}), which is affected by the size of our map pixels and is not fine-grained enough to separate out the multiple stellar populations present toward the GC.

\end{abstract}


\section{Introduction}
Interstellar extinction by dust has long been the bane of observers.
\cite{Trumpler1930} realized that there was interstellar extinction, and then estimated by comparing photometric distances (via the distance modulus) to geometric distances (derived after assuming clusters' spatial sizes were roughly linear with the number of stars and calculating the distance based on their derived angular size).
He concluded that some open clusters were much dimmer than expected due to astronomical extinction, and found that the extinction was consistent with a \SI{0.67}{mag} per \SI{}{\kilo\parsec}, given his assumptions.
The source of astronomical extinction is now largely thought to be due to the scattering of light off of dust grains in the interstellar medium (ISM) ~\citep[][and references therein]{draineLee1984, WeingartnerDraine2001, Foster2013}. 
 
 These dust grains are believed to be a mix of primarily silicates and carbon-based molecules (including graphite-based molecules and more complex polycyclic aromatic hydrocarbons, or PAHs), both with and without a veneer of frozen volatiles such as water or methane ices.
\cite{draineLee1984} modeled simple spherical silicate and graphite grains without an icy layer, which matched much of the observational data from optical to far-infrared wavelengths along many sightlines.
However, broad emission features in the mid-infrared (MIR) didn't fit simpler models; adding PAH molecules to the models brings them into better agreement with the data~\citep{WeingartnerDraine2001,DraineLi2007}.
In particular, broad emission features commonly attributed to PAH emission occur centered at MIR wavelengths near \num{3.3}, \num{6.2}, \num{7.7}, and \SI{8.6}{\micro\meter}~\citep{Leger1984PAH, Allamandola1989PAH, Rapacioli2005, Povich2007}, among other emission features.
The mechanism responsible is thought to be UV radiation reprocessed by PAH molecules and emitted at MIR wavelengths, principally via C$-$H bond stretching.
At near-infrared (NIR) and optical wavelengths Rayleigh scattering dominates, as the typical particle size is much larger than the wavelength ($R<<\lambda$)\footnote{
In the case of particles that are roughly the size of the wavelength of light ($R\sim\lambda)$, Mie theory is a more complete description of scattering.}.

\cite{CCM89} (henceforth CCM) were able to model extinction along several lines of sight to bright stars with a single-parameter fit:
\begin{equation}
A_\lambda = A_V\left(a_\lambda + b_\lambda \left(R_V\right)^{-1}\right)
\label{eqn:ccmExtinction}
\end{equation}
where $R_V$ is the slope of the extinction curve in the $V$ band, defined as 
\begin{equation}
R_V = \frac{A_V}{E(B-V)}
\label{eqn:colorExcessBV}
\end{equation}
 the ratio of the absolute extinction in $V$ band to the color excess; the commonly adopted value of $R_V$ in the Milky Way is 3.1~(CCM) with a range of \numrange{2}{5.5}~\citep{Foster2013}. 
In Equation~\ref{eqn:ccmExtinction}, $a_\lambda$ and $b_\lambda$ are piece-wise defined functions broken at \SI{0.9}{\micro\meter}.
They are high-order polynomials in the optical bands, simplifying to a power law in the infrared with an index, $\beta$, of about 1.61~\citep[CCM; see also][and references therein]{Mathis1990,Foster2013}.
We note others have found values of $\beta\sim2$ or more~\citep{indebetouw2005} particularly in the direction of the GC~\citep{Nishiyama2006,galacticnucleus2017,Hosek2018}.
The IR extinction equation is of the form
\begin{equation}
A_\lambda \propto \lambda^{-\beta}
\label{eqn:simpleIRextinction}
\end{equation}
where the wavelength range is approximately \SI{0.9}{\micro\meter} $<\lambda<$ \SI{10}{\micro\meter}.
We discuss infrared extinction laws in \S\ref{subsec:extinctionLaws} in more detail. 

While CCM found that, while on average the value of $R_V$ is 3.1, there is no reason \emph{a priori} that $R_V$ should be the same along all lines of sight due to the inhomogeneity of the Milky Way Galaxy disk structure.
\cite{WeingartnerDraine2001} interpret the value of $R_V$ as being linked to the physical size of the grains of dust, with larger grains having larger $R_V$ values.
This is an appealing model linking observations and theory.

The approach we have taken to measure the infrared extinction law at the Galactic Center is to use the Rayleigh-Jeans Color Excess method to determine the amount of extinction in map cells that are \SI{5}{\arcsecond} on a side.
We do this by creating surface brightness maps in both the H\footnote{
We use 2MASS filters for our near-infrared filter set, and cross-calibrate with the ISPI point source catalog, which in turn is well-calibrated to the 2MASS Point Source Catalog.}
and $[4.5\mu]$\footnote{
We use the \emph{Spitzer} IRAC filters for our mid-infrared colors, and denote them with [$x\mu$], where $x$ is the effective wavelength of each filter (\num{3.6}, \num{4.5}, \num{6.8}, and \SI{8.0}{\micro\meter)}, using the same convention as \cite{dustyVeilI,dustyVeilII,dustyVeilIII}.}
bands that are well-aligned to each other.
We generate an (H$-[4.5\mu]$) color map from these surface brightness maps and use it as an input in a Bayesian framework to measure the extinction law toward the Galactic Center.

In \S\ref{subsec:StellarColor} we have a brief discussion on filters and stellar color before introducing the Rayleigh-Jeans Color Excess (RJCE) method of estimating extinction in \S\ref{subsec:RJCE}.
Another method of measuring extinction, based on measuring Red Clump stars, is discussed in \S\ref{subsec:RC}.
We then delve into the assumptions behind NIR extinction laws in \S\ref{subsec:extinctionLaws}.
In \S\ref{sec:GCdata} we describe the datasets, first the NIR ISPI image data in \S\ref{subsec:ispiDataset}, followed by a discussion on the data reduction in \S\ref{subsec:ispiCalibration}.
Similarly, \S\ref{subsec:spitzerData} describes the \emph{Spitzer} MIR image data, while \S\ref{subsec:spitzerCalibration} describes how we reduced the \emph{Spitzer} data.
\S\ref{subsec:mapmaking} describes how we build our color maps.
In \S\ref{sec:InovationSurfBright} we discuss the innovative approach we have taken to produce extinction maps using surface brightness, color map, and the RJCE.
Then in \S\ref{sec:GCresults} we show our extinction map, compare our results to the literature, and apply our extinction map to create de-reddened color-magnitude diagrams derived from the ISPI point source catalog.
We conclude this work in \S\ref{sec:GCFutureWork} with a brief discussion on ideas for improvement and future work. 
We discuss our Bayesian approach, priors selection, and how it affects our results in Appendix~\ref{sec:GCPriors}.

\subsection{Stellar Color and Filter Selection}
\label{subsec:StellarColor}
\begin{figure}[h]
	\centering
	\begin{minipage}{0.48\textwidth}
		\includegraphics[width=\textwidth]{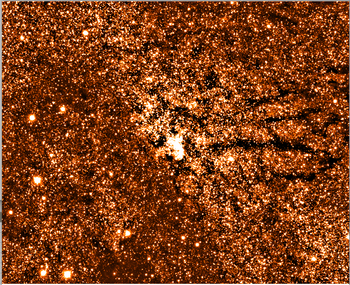}
	\end{minipage}
	\hfill
	\begin{minipage}{0.48\textwidth}
		\includegraphics[width=\textwidth]{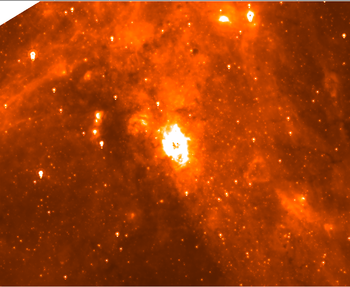}
	\end{minipage} 
	\caption{
		A zoomed-in region of the GC.
		The images are centered on the GC and about \SI{11}{\arcminute} by \SI{9.5}{\arcminute} in size.
		\textbf{Left:} The GC in H band.
		\textbf{Right:} The GC in the \emph{Spitzer} $[8.0\mu]$ band.
		Note the glowing, nebulous structure throughout the image; the PAH emission at \num{7.7} and \SI{8.6}{\micro\meter}, as well as warm dust, trace out complex morphology and is nonstellar in origin.
	}% eo caption
	\label{fig:emissionComparison}
\end{figure}

\begin{figure}[ht]
    \begin{minipage}{0.48\textwidth}
    	\includegraphics[width=0.95\textwidth]{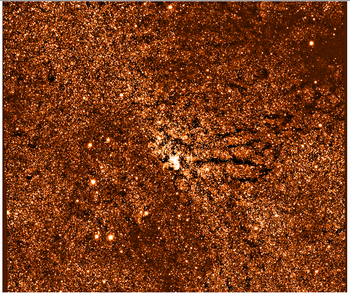}
	\end{minipage}
	\hfill
	\begin{minipage}{0.48\textwidth}
	        \includegraphics[trim=0 25 0 0, clip, width=\textwidth]{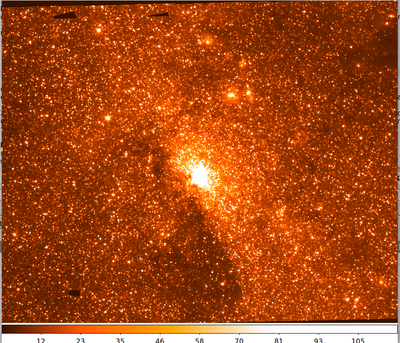}
	\end{minipage}
	\caption{As in Figure~\ref{fig:emissionComparison}, the images are centered on the GC.
	    North is up and East is to the left.
	    \textbf{Left:} The GC in H band.
	    The fully-reduced ISPI H band image of the GC.
	    North is up and East is to the left.
	    Note in particular the dark stripes running from right to left of the image (the `tiger stripes'); the lack of stars in the H band indicate a very high extinction in those regions.
	    The image is \SI{17}{\arcminute} by \SI{16}{\arcminute} (RA and Dec, respectively) and is centered on the GC.
	    \textbf{Right:} The \emph{Spitzer} $[4.5\mu]$ band image of the GC.
		Black regions are data reduction artifacts and have 0 counts.
		This image shows less evidence of extinction as there are no dark stripes to the right of the GC as there are in the H band image.
    	By eye several regions stand out as having an underdensity in stars, particularly the large region to the south and east and a slightly smaller region north and west of the GC.
	} % eo caption
		\label{fig:hBand}
        \label{fig:45Band}
\end{figure}

\begin{figure}[htb]
	\centering
	\includegraphics[width=0.95\textwidth]{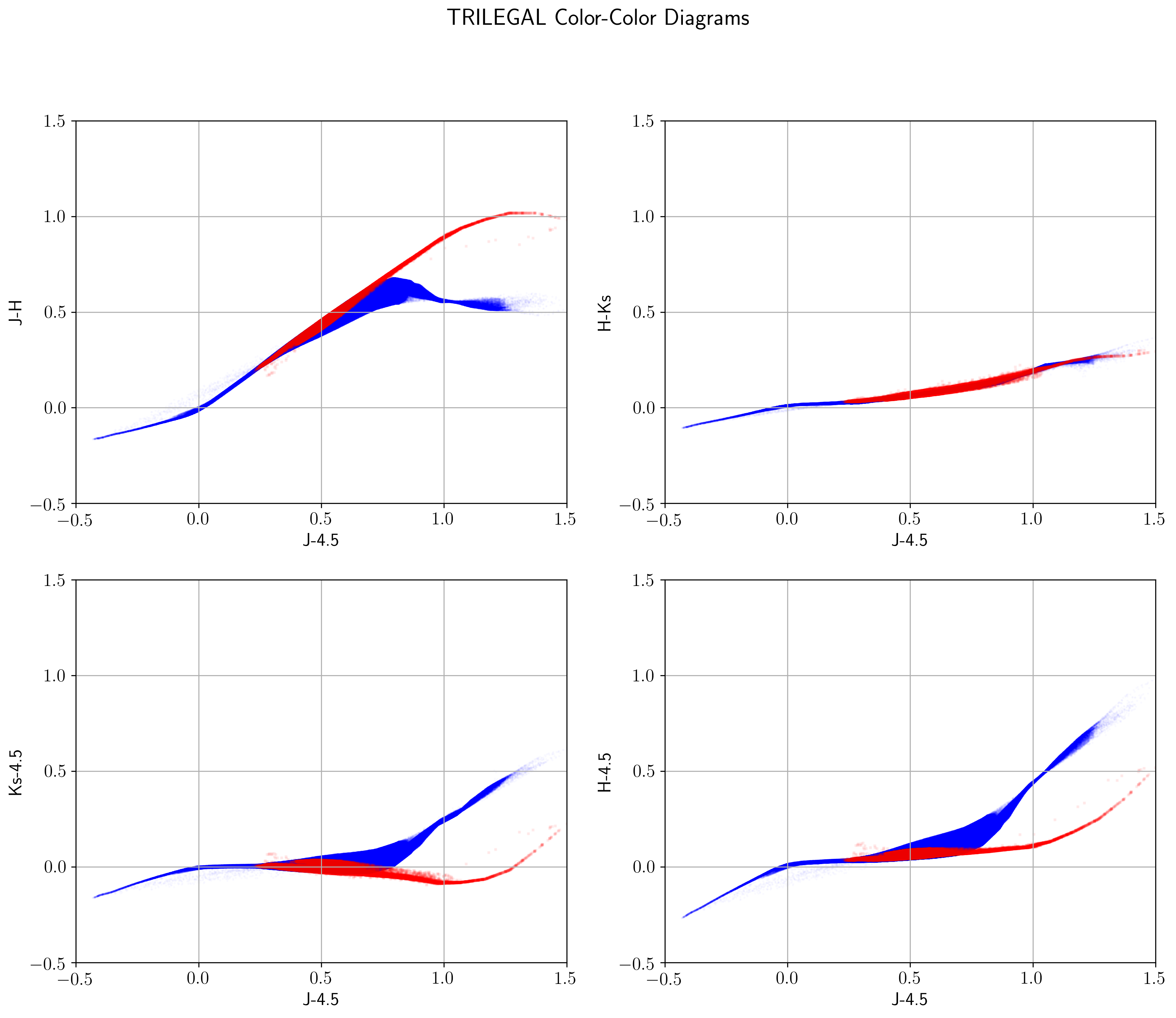}
	\caption{NIR-MIR color-color diagrams.
		Stars that have evolved off the main sequence are red and main-sequence stars are blue.
		Note the tight relation of both (K\textsubscript{S}-$[4.5\mu]$) and (H-$[4.5\mu]$). The overall scatter in color of all stars is smaller in (H-$[4.5\mu]$) than in (K\textsubscript{S}-$[4.5\mu]$).
	} %eo caption
	\label{fig:nirMirColorColorAll}
\end{figure}

Stars, to first order, are blackbodies, modulo metallicity, age, and mass which act to broaden the range of intrinsic colors.
Non-stellar emission from interstellar dust, such as PAH emission features centered at \num{3.3}, \num{6.2}, \num{7.7}, and \SI{8.6}{\micro\meter}~\citep{Leger1984PAH, Allamandola1989PAH, Rapacioli2005, Povich2007}  coincide in wavelength space with the $[3.6\mu]$, $[6.8\mu]$, and $[8.0\mu]$ \textit{Spitzer} filters.
Figure~\ref{fig:emissionComparison} shows side by side the H band and $[8.0\mu]$ images of the GC; the striking difference between the two images demonstrates the strongly non-stellar origin of PAH emission, especially in comparison to Figure~\ref{fig:45Band}, which shows the $[4.5\mu]$ band image next to the H band image.
The $[4.5\mu]$ filter the only \textit{Spitzer} filter left uncontaminated by PAH emission; we can and have assumed that the flux in this filter is largely  stellar blackbody emission.
We used the TRILEGAL model~\citep{trilegal2012} to generate several intrinsic color-color diagrams shown in Figure~\ref{fig:nirMirColorColorAll}.
The TRILEGAL model uses the~\cite{Girardi2002} isochrones for a wide range of ages ($7\leq\log{\textrm{(age/yr)}}\leq10.15$), masses ($0.08\leq M_\odot < 21$), and metallicities ($-1.5\leq\left[\textrm{Fe/H}\right]\leq0.18$), using a Chabrier IMF to generate three stellar populations representing the disk, bulge, and halo.
In Figure~\ref{fig:nirMirColorColorAll}  main-sequence stars are blue and evolved stars (RC stars, discussed in \S\ref{subsec:RC}, and RGB stars) are red.
Of note is the relatively tight distribution in both (K\textsubscript{S}$-[4.5\mu]$)$_0$ and (H$-[4.5\mu]$)$_0$ regardless of stellar type; this is particularly true for evolved stars.
The intrinsic spread (standard deviation) in color space when considering all stars is slightly lower in (H$-[4.5\mu]$)$_0$ compared to (K\textsubscript{S}$-[4.5\mu]$)$_0$, which motivates us to use (H$-[4.5\mu]$)$_0$ in this paper.

\subsection{Rayleigh-Jeans Color Excess Method}
\label{subsec:RJCE}
\begin{figure}[ht]
	\centering
	\includegraphics[width=0.95\textwidth]{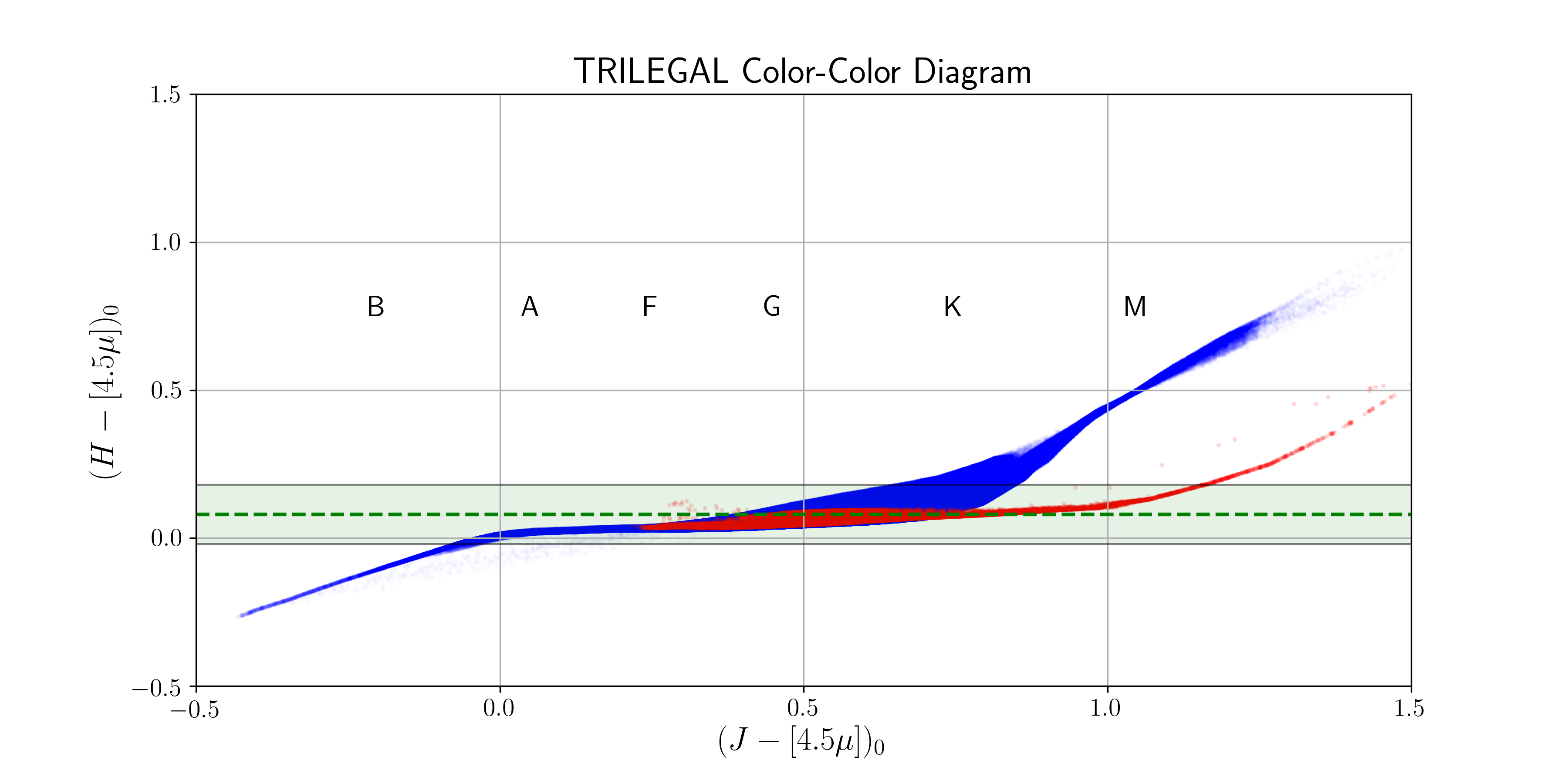}
	\caption{
		NIR-MIR color color diagram of the stellar population towards the Galactic Center.
	    Colors correspond to Main Sequence (in blue) and Red Giant Branch \& Red Clump stars (in red).
	    This plot shows the intrinsic color of stars with a wide range of ages ($7\leq\log{\textrm{(age/yr)}}\leq10.15$) and metallicities ($-1.5\leq\left[\textrm{Fe/H}\right]\leq0.19$), taken from the Padova-Girardi isochrones~\citep{Girardi2002} and generated using the TRILEGAL model~\citep{trilegal2012} by setting the extinction to 0.
	    The green dotted line marks the color locus (H$-[4.5\mu]$)$_0=0.08$ and the shaded light green box marks $\pm$\SI{0.1}{mag}.
	} %eo caption
	\label{fig:nirMirColorColor}
\end{figure}

The RJCE method~\citep{dustyVeilI,dustyVeilII,dustyVeilIII} hinges on the fact that stars have a fairly uniform intrinsic color in the infrared.
This allows us to assume that all stars have nearly identical color for a certain set of filters.
For example, nearly all stars have an identical color (H-[$4.5\mu$])$_0$, which is calculated by subtracting the magnitude in IRAC Channel 3 band (effective wavelength: \SI{4.442}{\micro\meter}\footnote{
The effective wavelengths quoted above are calculated by convolving the SED of a K2 III star with the each of the filters; using a flat-spectrum (e.g., no weighting) results in small shifts in the effective wavelength~\citep[see footnote 13 of][]{indebetouw2005}.}
) from H band (effective wavelength: \SI{1.664}{\micro\meter}), regardless of their age, metallicity, mass, and a host of other physical properties.
Figure~\ref{fig:nirMirColorColor} shows an (H$-[4.5\mu]$)$_0$ vs (J$-[4.5\mu]$)$_0$ color-color diagram generated via the TRILEGAL model in order to derive the intrinsic (H$-[4.5\mu]$)$_0$ color scatter across the wide swathe of model stellar type and age.
Earlier work by~\cite{dustyVeilI} found that the intrinsic color, (H-[$4.5\mu$])$_0$, is about \SI{0.08}{mag} for a wide range of stars, with a scatter of about \SI{0.1}{mag} over F, G, and K stars.

We found that the scatter in (H$-[4.5\mu]$)$_0$ is at most \SI{0.4}{mag} across all stellar types (B-M); if one excludes the  O, B, and M dwarfs, we find that the scatter over A, F, G, and K stars from the TRILEGAL model is about \SI{0.1}{mag}, in good agreement with~\cite{dustyVeilI}.
This deliberate exclusion of O, B, and M dwarfs is motivated by the fact that similarly, the hottest O and B dwarfs are the rarest stars in an imaging survey.
Similarly, M dwarfs, while the most plentiful stars in the Galaxy, are also intrinsically the dimmest and therefore are insignificant in our ISPI and \emph{Spitzer} images.
\label{para:nirMirColorColor}
Additionally, our technique averages over a number of stars in each map pixel, and this dilutes the effects of a single O, B, or M dwarf star in a pixel.

Other colors were considered by ~\cite{dustyVeilI}, but were discarded in favor of (H$-[4.5\mu]$).
A potential alternative color is (K\textsubscript{S}$-[4.5\mu]$), which exhibits a smaller scatter than (H$-[4.5\mu]$). 
We opted to not use K\textsubscript{S} in order to preserve a more direct connection with~\cite{dustyVeilI}.
Also, in terms of spectroscopic follow-up measurements of extinction along the sight-lines of single stars, the APOGEE survey is a natural choice, but it is limited to H-band spectroscopy, and this makes using H-band photometry more appealing.
Further, we note that at the GC, the contamination from PAH emission features at \num{3.3}, \num{6.2}, \num{7.7}, and \SI{8.6}{\micro\meter} is severe enough that only the $[4.5\mu]$ MIR images are useful for determining color using starlight.

\subsection{The Red Clump Extinction Method}
\label{subsec:RC}
A popular technique for estimating interstellar extinction using only NIR PSF photometry is the Red Clump (RC) method to estimate both extinction power law exponents and extinction values~\citep{Nishiyama2006, galacticnucleus2017,Hosek2018}.

We note that this method is solidly independent of our surface brightness approach.
The RC method can be summed up thusly: If one assumes that RC stars all exhibit the same NIR colors, one can infer the amount of extinction based on the amount of reddening in the three NIR bands.
This method assumes that because RC stars have just begun their helium-core burning phase and thus are producing the same amount of energy regardless of total stellar mass, core mass, or metallicity (and exhibit a small intrinsic color range as a result,~\cite{salarisgirardi2002})  that they can be considered standard candles.
With many thousands of color excess measurements of RC stars, one can build up a reddening law as well.
However, by using only NIR bands and restricting the stars used to those with measurable J-band magnitudes, this method necessarily probes a shallower distance toward the GC than our surface photometry simply due to the extreme reddening present in the direction of the GC.
We compare our results to other RC measurements of the NIR extinction law in \S\ref{sec:GCresults}.

\subsection{Extinction Laws in the Infrared}
\label{subsec:extinctionLaws}
Historically, extinction in the Milky Way Galaxy has been measured in the optical along many lines of sight and then fit to a single parameter, $R_V$ from Equations~\ref{eqn:ccmExtinction} and \ref{eqn:colorExcessBV}~\citep{CCM89,dustyVeilI,dustyVeilII,dustyVeilIII,Foster2013}.
In the near-infrared, extinction laws are well-fit with a simple power law, which can be written as 
\begin{equation}
\log_{10}\left[{\frac{A_\lambda}{A_{K_S}}}\right] = C - \beta\log_{10}\left[\lambda\right]
\label{eqn:simpleExtinct}
\end{equation}
 where $\beta$ is the slope of the extinction fit as a function of $\lambda$, and $\lambda$ is the effective wavelength of the filters used to make the measurements~\citep{indebetouw2005}.
The constant, $C$, is the value of the equation at the limit $\lambda\rightarrow\infty$; the `standard' way of finding $C$ is to extrapolate the relative extinction values and is found to be about 0.6~\citep{indebetouw2005,Nishiyama2006}.
The relative extinction, $\frac{A_\lambda}{A_{K_s}}$, is found by measuring the color excess relative to K\textsubscript{S} and normalizing to one value of relative extinction (in other words, making an assumption, or measurement, of the extinction law at two wavelengths)~\citep{Rieke1985}. 
\cite{indebetouw2005} found that when incorporating mid-infrared measurements of Red Clump stars (see \S\ref{subsec:RC}), the infrared extinction is better fit with a second-order polynomial in log-space:
\begin{equation}
\log_{10}\left[{\frac{A_\lambda}{A_{K_S}}}\right] = C - \beta_1\log_{10}\left[\lambda\right] + \beta_2\left(\log_{10}\left[\lambda\right]\right)^2
\label{eqn:ExtincSquared}
\end{equation}
where $\beta_2$, $\beta_1$, and $C$ are fit using a weighted least-squares approach, and the $\log_{10}[\lambda]$ terms are calculated by using the effective wavelength (in microns) of the filter in question. %CONSIDER REMOVING THIS PARENTHETICAL (e.g., the 2MASS H band filter has an effective wavelength of \SI{1.664}{\micro\meter}, and the \emph{Spitzer} $[4.5\mu]$ band \SI{4.442}{\micro\meter}).	
We adopted a fixed value of  $C=0.60$.
In contrast,~\cite{Nishiyama2006} using the RC method found that $\beta=1.99\pm0.02$ and $C=0.494\pm0.004$.

We use (H $- [4.5\mu]$) to calculate the extinction in K\textsubscript{S} (or A(K\textsubscript{S})).
Calculating the relationship between the color (H$-[4.5\mu]$), $\beta$, and A(K\textsubscript{S}) is straightforward.
We begin with two equations,  $\left(\log_{10}\left[\frac{A_H}{A_{K_s}}\right]\right)$ and $\left(\log_{10}\left[\frac{A_{[4.5\mu]}}{A_{K_s}}\right]\right)$ and note that color excess normalized by the extinction in K\textsubscript{S}, $\frac{E(\textrm{H}-[4.5\mu])}{A(K_s)}$, is the same as $\left(\frac{A_H}{A_{K_s}}-\frac{A_{[4.5\mu]}}{A_{K_s}}\right)$.
With some algebra, we can reduce the two equations to 
\begin{math}
\frac{E(\textrm{H}-[4.5\mu])-\left(\textrm{H}-[4.5\mu]\right)_0}{A_{Ks}} =  10^{C-\beta\log_{10}\lambda_\textrm{H}} - 10^{C-\beta\log_{10}\lambda_{[4.5\mu]}}
\end{math}
In this equation, $\lambda$\textsubscript{H} and $\lambda_{[4.5\mu]}$ are simply the effective wavelength in \SI{}{\micro\meter} of the filters (1.664 for H, 4.442 for $[4.5\mu]$).
The intrinsic color, (H$-[4.5\mu]$)$_0$, must also be be subtracted from the color excess; its value is \SI{0.08}{mag}~\citep{dustyVeilI}.
Solving for $A_{K_s}$, we find
\begin{eqnarray}
A_{K_s} = \frac{E(\textrm{H}-[4.5\mu])-0.08}{10^{C-\beta*\log_{10}\lambda_\textrm{H}}-10^{C-\beta\log_{10}\lambda_{[4.5\mu]}}}
\label{eqn:IRextinction}
\end{eqnarray}

With our data we are able to fit $\beta$ as well as measure the extinction.
The RJCE method thus requires only a measured (H$-[4.5\mu]$) value and Equation~\ref{eqn:IRextinction} with a suitable $\gamma$:
\begin{equation}
A_{K_s} = \gamma \left[\textrm{E}(\textrm{H}-[4.5\mu]) - 0.08\right]
\label{eqn:dustyVeilExtinction}
\end{equation}
where $\gamma = 1/(10^{C-\beta*\log_{10}\lambda_\textrm{H}}-10^{C-\beta\log_{10}\lambda_{[4.5\mu]}})$.

Plugging in the values for $C$ and $\beta$ from~\cite{indebetouw2005} and and using our effective wavelengths, we calculate $\gamma$ to be 0.918, in agreement with~\cite{dustyVeilI,dustyVeilII,dustyVeilIII}.
We used the simpler extinction power law for our work and found that we were able to independently fit for $\beta$ using our Bayesian approach, described in more detail in Appendix~\ref{sec:GCPriors}.

\section{The Dataset}
\label{sec:GCdata}
We use imaging data of the GC taken with the Infrared SidePort Imager (ISPI) camera on the Cerro-Tololo International Observatory \SI{4}{\meter} Blanco Telescope~\citep{ispi2004} in 2005, for our H-band photometry.
This dataset was obtained and used by \cite{dewitt2010} \nocite{dewittThesis2011} to find infrared counterparts to X-Ray sources.
\cite{dewitt2010} generated a point-source catalog using DAOPHOT~\citep{stetson1987}, first reducing the images using the Florida Analysis Tool Born Of Yearning for high-quality astronomical data (FATBOY) and then cross-checking their point-source catalog against the 2MASS point source catalog~\citep{2mass2006}.
We used the well-calibrated ISPI Point Source Catalog (henceforth IPSC) created by \cite{dewitt2010} as ground-truth for our ISPI data reduction and calibration.
We re-reduced the ISPI images for this work using a revamped version of FATBOY that is optimized for GPUs named superFATBOY~\citep{prelim-sFB2012}.

For the $[4.5\mu]$ data we used the InfraRed Array Camera (IRAC)~\citep{fazio2004} onboard the \emph{Spitzer Space Telescope} taken for the Galactic Legacy Infrared Midplane Survey Extraordinaire, or GLIMPSE, survey~\citep{glimpse2003, glimpse2009}.
These data were downloaded using the website for the NASA Infared Processing and Analysis Center (IPAC) Infrared Science Archive (IRSA), hosted at the Jet Propulsion Laboratory at the California Institute of Technology\footnote{The GLIMPSE website: \\ \href{http://irsa.ipac.caltech.edu/data/SPITZER/GLIMPSE/overview.html}{http://irsa.ipac.caltech.edu/data/SPITZER/GLIMPSE}}.
We selected the \SI{0.6}{\arcsecond} images to assist with reducing the number of sources that are saturated.

\subsection{The ISPI Dataset}
\label{subsec:ispiDataset}
The ISPI camera has a \SI{10.5}{\arcminute} square field of view with a plate scale of \SI{0.3}{\arcsecond\per pixel}~\citep{ispi2004}.
GC data were obtained in J, H, and K\textsubscript{S} bands on 10 August 2005.
Four fields were observed, each using a 4-point dither pattern with a dither step size of \SI{20}{\arcsecond} to cover a \SI{10.3}{\arcminute} field and to remove the detector's cosmetic defects; the four overlapping fields were tiled to make a final field of view of \SI{17}{\arcminute}$\times$\SI{17}{\arcminute}.\footnote{
To try and reduce the confusion between the four fields and the 4-point dither pattern, we spell out numbers (e.g., one, four) when discussing fields and use numerals (e.g., 1, 4) when discussing the dither pattern.}
The centers of the four fields make a square \SI{420}{\arcsecond} on a side centered on the GC.
After the dither sequences were completed for all four fields, the telescope was re-centered on the GC to check for focus; as the focus stability was very good and required no adjustment, we effectively have an additional field from this data, albeit with a different exposure time than the four `quadrant' fields.

The individual frame exposure times in J, H, and K\textsubscript{S} were \SI{5}{\second}, \SI{3.2}{\second}, and \SI{3.2}{\second}, respectively.
The frame exposure times were kept short to prevent saturation of the brightest stars in the field, or at least minimize the number of saturated pixels.
Each of the four fields had a total exposure time of \SI{200}{\second}, \SI{113}{\second}, and \SI{32}{\second} for J, H, and K\textsubscript{S}, respectively.
The center field, made with the `focus check' frames, had a total of \SI{60}{\second}, \SI{68}{\second}, and \SI{68}{\second} for J, H, and K\textsubscript{S}, respectively.
We then made a master image out of all five fields, giving a rather deep K\textsubscript{S} image of the center of the image in comparison to the corners.

After the 4-point dither sequence was completed at each field, a less-crowded off-source field about \SI{2}{\degree} away was observed for sky background estimation.
This off-sky field contained fewer bright sources than the GC field while still being close enough to give acceptable sky background estimates.
After reducing the ISPI data with superFATBOY, we produced master images in J, H, and K\textsubscript{S} roughly \SI{17}{\arcminute} by \SI{17}{\arcminute} in size.
The depth of this master image varies across  the \SI{17}{\arcminute} field of view and bandpasses, and ranges from \SIrange{200}{468}{\second}, \SIrange{113}{294}{\second}, and \SIrange{32}{132}{\second} for J, H, and K\textsubscript{S}, respectively.
The exposure time is deepest at the center of the field and shallowest at the corners.

\subsection{ISPI Calibration}
\label{subsec:ispiCalibration}

As noted above, we used the superFATBOY data reduction code~\citep{prelim-sFB2012} to reduce the ISPI data.
superFATBOY is written in Python and is massively parallelized to take advantage of NVIDIA's Compute Unified Device Architecture (CUDA) Graphics Processing Unit (GPU) devices with hundreds to thousands of GPU cores.\footnote{
superFATBOY can also be run in CPU-mode, albeit with considerably slower run time in comparison to GPU-mode.}
It is based on the CPU-only FATBOY-SLIM (FATBOY-Sans Lousy IRAF Mistakes) Python code with modifications geared toward making superFATBOY usable for any infrared or optical imaging or spectroscopy while depending on stable Python scientific packages like \texttt{astropy} and \texttt{scipy}. 
The algorithms used by superFATBOY are not instrument-specific (although  modules that are instrument-driven can be incorporated) to reduce infrared data follow basic standards, such as distortion correction, dark-subtraction, flat-fielding, bad pixel masking, sky background estimation \& subtraction, and image stacking.

Sky background estimation \& subtraction is one of the more delicate operations when reducing GC data.
Because the infrared sky background levels change drastically in a stochastic manner over timescales of minutes, background estimation can be fraught even for uncrowded fields.
In the GC, the source density is so high that we cannot get a `clean' estimate of the sky background by using the scientific data.

The `off-GC' field used for sky background estimates contained fewer bright sources than the GC field while still being close enough to give decent sky background estimates.

The ISPI data were reduced using superFATBOY's off-source sky subtraction method.
We made several iterations to find which sky background interpolation method worked best\footnote{We used the ISPI dataset as a testbed for the alpha version of superFATBOY; when results were in agreement with the previous data reduction performed by~\cite{dewittThesis2011}, we declared superFATBOY a success.}.
In the end we ended up using a 2-pass sky estimator algorithm to better remove the sources present in the off-source skies, and used the `nebular' flavor of scaling the background to the median of the science frames.
Additionally we restricted the off-source frames that went into making the master sky background frames to be drawn from those off-source frames taken nearest in time to the science frames; in other words, each dither sequence used the off-source frames taken immediately after or before the sequence was completed.
Ordinarily all the off-source sky frames go into making the master sky background image, but because the observations were distributed in time over the course of about an hour, the sky background could not be considered constant.

Once we had the reduced master images, we corrected and made uniform the astrometry of each image.
In order to have a uniform astrometric solution across all of our images, we used a 3 step process applied to each of the master images.
By having a uniform astrometric solution, we were able to resample all of the images onto a common pixel grid.
This is motivated by our desire to make color maps, which require the parent images to have the same pixel grid as much as is possible.
We used Source Extractor~\citep{sextractor1996}, Scamp~\citep{scamp2006}, and SWarp~\citep{swarp2002} to correct the astrometry to a common pixel grid.
The algorithm we used for astrometric correction is as follows:
	\begin{enumerate}
	\item \textbf{Source Extractor}
	  Find the pixel coordinates of stars (we restricted Source Extractor to use bright but unsaturated stars) and record the plate scale of the image in a catalog.
	\item \textbf{Scamp}
	  Compare the pattern of the stars' locations from the Source Extractor catalog and, using an initial guess of the location on the sky provided by the FITS header, use the 2MASS point source catalog to calculate both new astrometry and map any residual distortion present in the image.
	\item \textbf{SWarp}
	  Apply the astrometric and distortion correction from Scamp.
\end{enumerate}
SWarp also provides the ability to set the output pixel scale.
It uses a flux-conserving bi-cubic interpolation algorithm to resample the images onto the new grid.
We used this function to make our output pixels \SI{0.3}{\arcsecond}, or unchanged from the ISPI raw frame plate scale; we did this to simplify the map-making step later on.
This process was repeated with the \emph{Spitzer} data as described in \S\ref{subsec:spitzerCalibration}.

With our reduced and astrometrically-corrected images, we were able to make direct comparisons with the IPSC~\citep{dewitt2010} and use this catalog to find the zero-point (ZP) magnitude of our newly-reduced images and calibrate the photometry.
First we made a Source Extractor catalog of the stars in our images.
We then used the stars in our catalog with a `signal/noise' ratio $>500$ (where `signal' and `noise' were taken from the Source Extractor columns `FluxBest' and `errFluxBest', respectively) and, using the RA and Dec coordinates from our catalog, matched them to stars in the IPSC within \SI{1}{\arcsecond} on the sky.
Of the 749 stars that were selected with a signal/noise ratio $>500$, 187 were rejected as having no counterpart in the IPSC.
A further 15 were rejected for having two counterparts within \SI{1}{\arcsecond}, leaving a total of 547 singly-matched stars.
We examined the curve of growth for singly-matched stars to find the aperture size which appeared to capture all or most of the flux of each star and found that a \SI{2}{\arcsecond} aperture is where the curve of growth flattened out.

With the aperture set, we then found the instrumental magnitude for each star in our catalog by taking the log of the total flux in the \SI{2}{\arcsecond} aperture and multiplying by -2.5:
\begin{equation}
\textrm{instrumental magnitude} = -2.5\log_{10}(\textrm{total flux})
\end{equation}
To find the zero-point, we then took the magnitude from the IPSC and subtracted our instrumental magnitude:
\begin{equation}
\textrm{zero point} = \textrm{IPSC magnitude} - \textrm{instrumental magnitude}
\end{equation}
We found that our instrumental ZP magnitude in H-band is best fit with a Gaussian with a mean of \SI{22.26}{mag} and a standard deviation of \SI{0.09}{mag}.
We do not quote the reduced error ($N^{1/2}=(547)^{1/2}$\SI{0.004}{mag}) in order to be more conservative in our error estimation of our photometry.

Estimating the background at the GC is extremely difficult due to the high stellar density. 
Typically one uses an annulus centered around each point source with an inner radius of \SIrange{10}{20}{\arcsecond} away, but at the GC, the crowding limit makes this approach worse than useless as each line of sight typically ends in another (albeit unresolved) star or stars.
We do not perform background subtraction for the stars used to perform our ZP measurement because:
\begin{enumerate}
	\item The stars we have selected for ZP measurement range in magnitude from H=10.5 to H=12.
	  These stars are bright enough that the contribution from the background is not the dominant source of error.
	  They are not bright enough to have saturated the detector, nor are they bright enough to be in the non-linear regime of the detector.
	\item The estimated magnitude errors for these stars in the IPSC is on average 0.014 magnitudes with a standard deviation of 0.013 magnitudes.
\end{enumerate}
We adopt the sum in quadrature of the standard deviation to our zero-point fit and the average error from the IPSC stars as our estimated error in our photometry: $\sigma_{phott} = \sqrt{0.09^2+0.014^2}=0.091\approx0.09$.

\subsection{The \emph{Spitzer} Dataset}
\label{subsec:spitzerData}
We selected images from the GLIMPSE survey that covered as much of the ISPI field of view as possible.
In the interest of time and given that we were limited to image stamps smaller than \SI{600}{\arcsecond} on a side, we downloaded a mosaic belonging to dataset Spitzer\#\num{0013368832}.
This mosaic was created using 225 subimages of the GC and is \SI{77.5}{\arcminute} by \SI{26.8}{\arcminute} (Galactic longitude and latitude, respectively).
However, at the Galactic longitude we care about ($|l|<0.25$), this mosaic is only about \SI{16}{\arcminute} high in latitude, meaning that we are unable to generate colors involving any of the mid-infrared bands for about \SI{30}{\arcsecond} from the top and bottom of our ISPI field.

\subsection{\emph{Spitzer} Calibration}
\label{subsec:spitzerCalibration}
The\emph{Spitzer} images we downloaded were mosaicked and calibrated as part of the GLIMPSE survey.
In order to guarantee excellent astrometric matching between our ISPI and \emph{Spitzer} data, we used the Source Extractor $\rightarrow$ Scamp $\rightarrow$ SWarp algorithm described in \S\ref{subsec:ispiCalibration}, using Source Extractor to pull out bright but unsaturated stars from the \emph{Spitzer} images.
Instead of using the 2MASS catalog to correct the \emph{Spitzer} astrometry (as we did for the ISPI image), we used a new star catalog created from running Source Extractor on the fully reduced and Scamped/SWarped ISPI H-band image.
We did this for two reasons.
First, this allowed us to make a magnitude cut on the ISPI H-band Source Extractor catalog, our reasoning being that moderately bright stars in H-band should be reasonably bright at MIR wavelengths, without having to rely on the 2MASS point source catalog (which is not as deep as our H-band image).
Second, it also forced the \emph{Spitzer} images to match the astrometry of the H-band image after we performed the Scamp $\rightarrow$ SWarp algorithm.
We made the output plate scale match the ISPI plate scale (\SI{0.3}{\arcsecond \per pixel}), letting SWarp perform the re-sampling using its default (flux-conserving) settings.

The IRAC image values are in \SI{}{\mega\Jansky\per\steradian}.
To convert to Vega magnitudes, we simply use the following formula taken from Equation 4.19 in~\cite{IRAC2015}:
\begin{equation}
m_{\textrm{Vega}} = -2.5 \log_{10} (\textrm{flux}) + 2.5 \log_{10}(\textrm{ZP}/\textrm{C})
\label{eqn:ZPcorrection}
\end{equation}
where the zero-point (ZP) depends on the IRAC channel (\cite[Table 4.9]{IRAC2015}).
Unlike the ZP for ISPI, IRAC's ZP is the flux of a 0\textsuperscript{th} magnitude star in the Vega system; for [$4.5\mu$], $C=179.9\pm2.6$\SI{}{\Jansky}.

The correction factor, C, in Equation~\ref{eqn:ZPcorrection} is a conversion factor from \SI{}{\mega\Jansky\per\steradian} to \SI{}{\Jansky\per pixel}.
For the \SI{0.6}{\arcsecond} by \SI{0.6}{\arcsecond} pixels, C = \SI{8.461595e-6}{\Jansky\per pixel \per(\mega\Jansky\per\steradian)}.
For the Scamped/SWarped image with \SI{0.3}{\arcsecond} by \SI{0.3}{\arcsecond} pixels, C = \SI{23.5045e-6}{\Jansky\per pixel \per(\mega\Jansky\per\steradian)}.

We need to perform an aperture correction on the surface brightness measured at each pixel.
The \emph{Spitzer} Handbook \S4.11.3 gives the \emph{Spitzer} surface brightness corrections (for [$4.5\mu$], the fudge factor is 0.94) and at the end of section notes that the correction factors should be good to 10\%.
In other words, the `real' surface brightness SB can be calculated by 
\begin{equation}
\textrm{SB} = \textrm{sb}*\textrm{f}
\end{equation}
where sb is the measured surface brightness in a pixel and f is the correction fudge factor, which has the accuracy of about 0.1 magnitudes -- comparable to the ISPI photometric (zero-point) error, and, for the [3.6$\mu$] and [$4.5\mu$], $f$ is not significantly different from unity.
The correction factor for [4.5$\mu$] is \num{6}\%\, $\pm \sim$\num{10}\%, and when these are added in quadrature the result is about \num{10}\%.
As a result, we have not applied the correction to our [$4.5\mu$] photometry and instead adopt a blanket 10\% error for our surface brightness photometry error term.

\subsection{Map-Making}
\label{subsec:mapmaking}
The next part of our analysis required us to build an [H$-4.5\mu$] color map from the astrometrically-matched ISPI and \emph{Spitzer} images.
We used a map-making code written in Python by former UF graduate student Daniel Gettings.
The \texttt{map\_tools} module was designed to create two-dimensional histograms with the bin size chosen by the user while maintaining full World Coordinate System (WCS) information.
We generated maps with `pixels' or `cells' ranging in size from \SIrange{5}{60}{\arcsecond}, effectively summing the pixel values in each image that fall in each map's cell and converting that to a surface brightness in Vega magnitudes.
We note that \texttt{map\_tools} does conserve flux geometrically, allotting the fractional value to map cells of pixels that partially fall on those cells.

We settled on using the \SI{5}{\arcsecond} cell size for our maps because we wanted to use the smallest cells possible without partially resolving individual bright stars; given that the point spread function (PSF) of the IRAC $[4.5\mu]$ bandpass has a full-width at half-maximum (FWHM) of about \SI{1.1}{\arcsecond}, a \SI{5}{\arcsecond} $\times$ \SI{5}{\arcsecond} square aperture is about as small as we can go without single stars dominating a cell. 

Pixels from the original images that straddle cell borders are a potential source of `smearing' in the sense that the PSFs of bright stars may fall across cell boundaries and `contaminate' the cell with perhaps a few pixels' worth of light.
However, given that the cells have an area of \SI{25}{\arcsecond}$^2$, the contribution from a `contaminated pixel' is strongly diluted as the native pixel area is only \SI{0.09}{\arcsecond}$^2$, a factor of almost 300 smaller than the map cells.
Additionally, this `contaminated' light is not noise as it is still stellar in origin.
In the case of a star centered more or less on a cell border, we note that the intensity measured in the map cells reflect the fact that there is a star whose PSF falls in all cells it touches.

Once we generated our binned maps, creating color maps (e.g., H$-$K\textsubscript{S} or H$-[4.5\mu]$) was easy because all of the maps have identical cell and WCS coordinate grids.

\subsection{The Innovation: Surface Brightness}
\label{sec:InovationSurfBright}
Here we describe the RJCE method as we apply it to the GC region.
We use the binned maps made as described in \S\ref{subsec:mapmaking} to generate an (H$-[4.5\mu]$) color surface brightness map.
This map acts as our measurement for each \SI{5}{\arcsecond}$\times$\SI{5}{\arcsecond} cell.
In aggregate, stars in any one particular cell have a (H$-[4.5\mu]$)\textsubscript{0} color of $0.08\pm\sim0.1$; the errors originate in the intrinsic distribution of stellar color as shown in Figure~\ref{fig:nirMirColorColor} and discussed in \S\ref{subsec:RJCE}.
Thus, if we sum the colors of all sources (assumed to be stars), we can simply appropriate the intrinsic color and associated error as the expected color and color error of each cell.
This allows us to take the measured color excess and assign an extinction value to each cell.
Additionally, given the large amount of data (over \num{45000} map cells), we can use a Bayesian approach to find values of $\beta$ and the extinction A(K\textsubscript{S}) given a distribution of extinction laws derived from varying $\beta$ and the observed color excess.

\section{Results}
\label{sec:GCresults}
\begin{figure}[hbt]
    \centering
    \begin{minipage}{0.4\textwidth}
	    \centering
		\includegraphics[width=\textwidth]{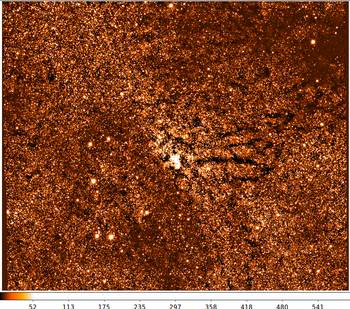}
	\end{minipage}
~~
	\begin{minipage}{0.4\textwidth}
		\centering
		\includegraphics[width=\textwidth]{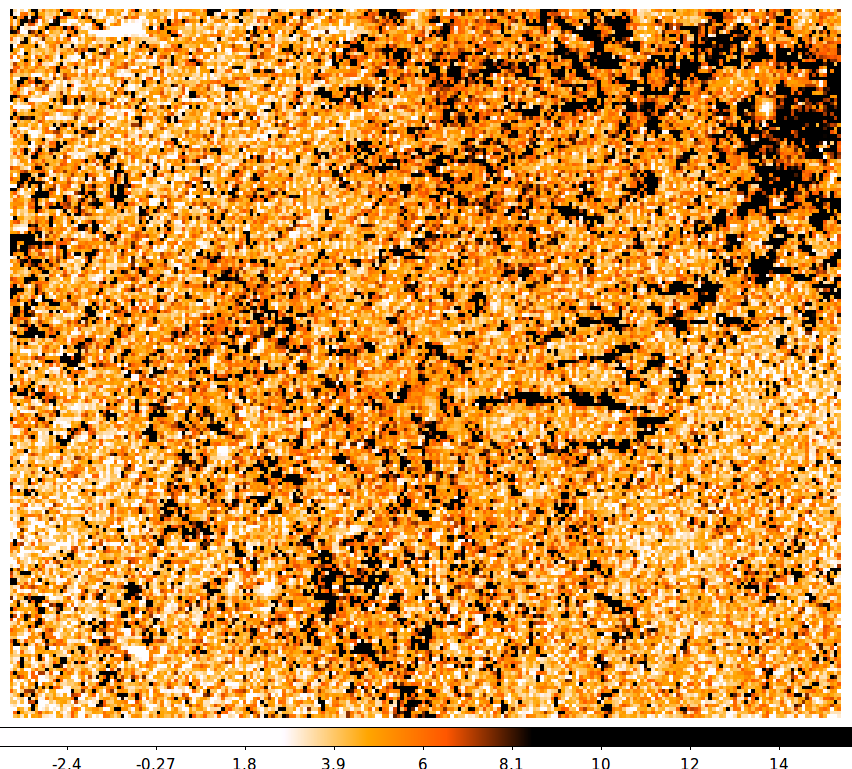}
	\end{minipage}
\\
    \begin{minipage}{0.4\textwidth}
		\centering
		\includegraphics[width=\textwidth]{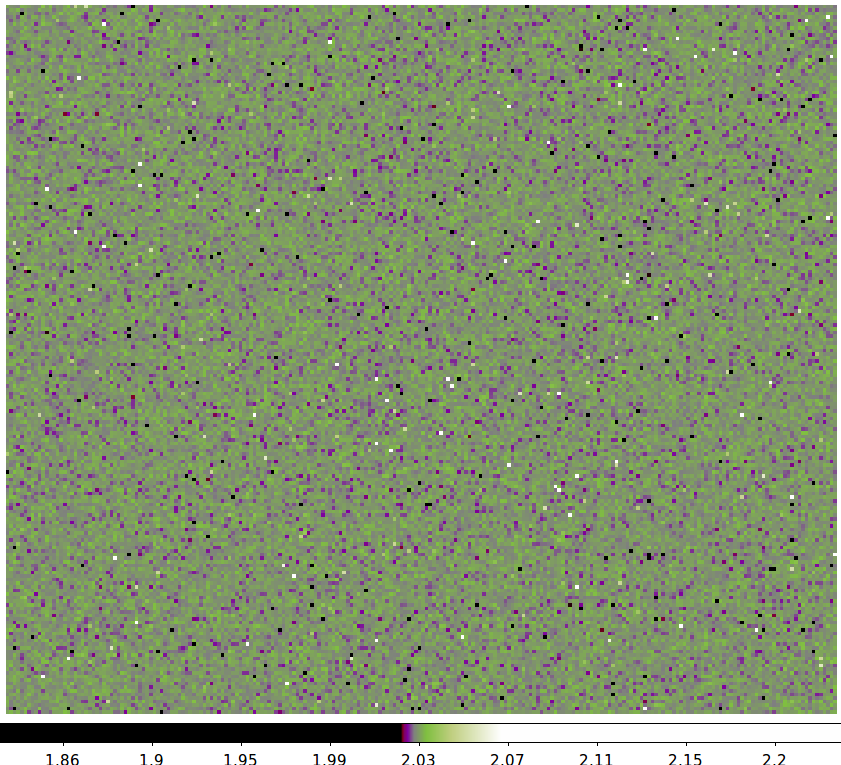}
	\end{minipage}
~~
\begin{minipage}{0.4\textwidth}
		\centering
		\includegraphics[width=\textwidth]{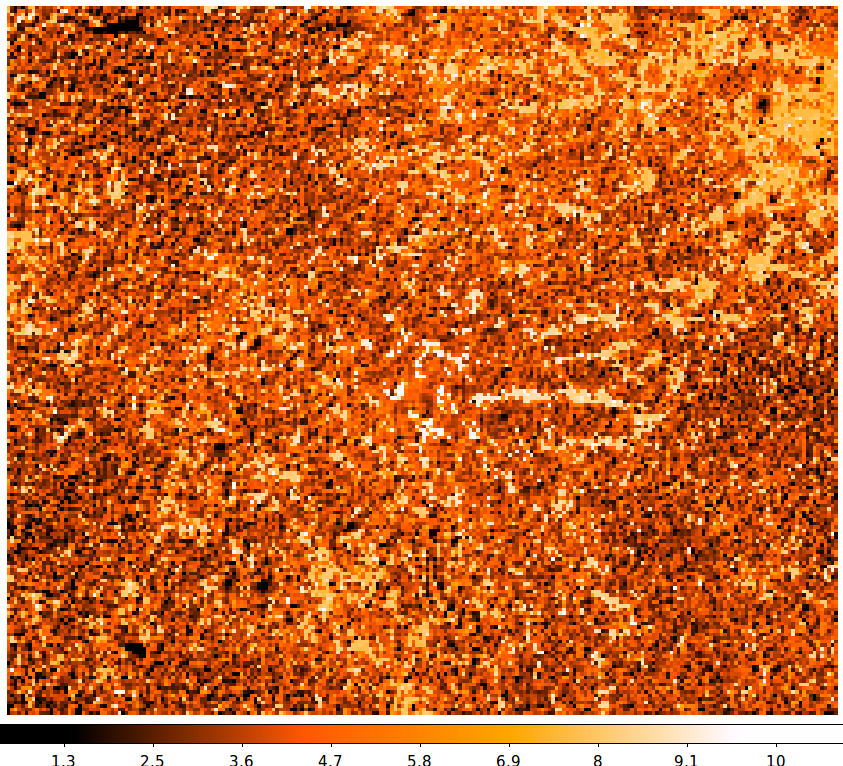}
	\end{minipage} 
\\	
	\begin{minipage}{0.4\textwidth}
		\centering
		\includegraphics[width=\textwidth]{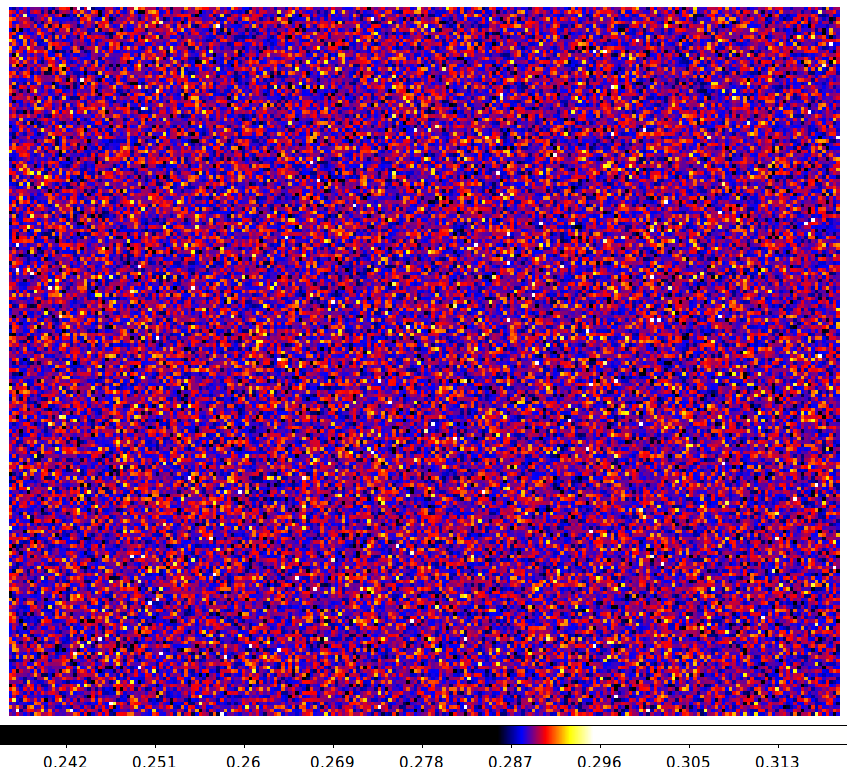}
	\end{minipage}
~~
\begin{minipage}{0.4\textwidth}
		\centering
		\includegraphics[width=\textwidth]{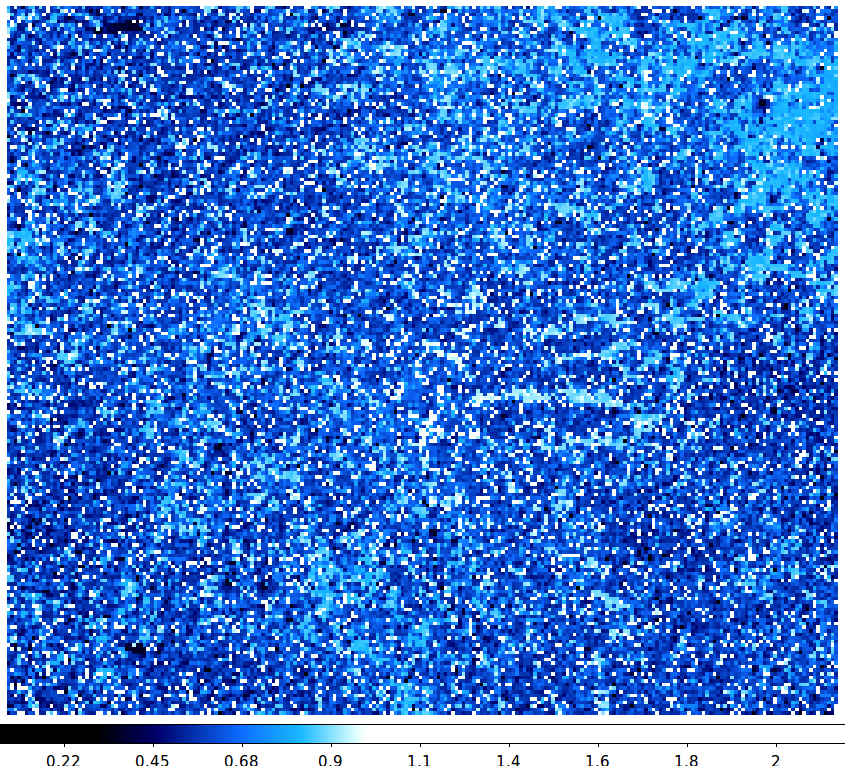}
	\end{minipage}
    	\caption{
    	Maps of the GC.
		All of the images have been aligned to show the same field.
		\textbf{Top left:} The GC in H band.
		\textbf{Top right:} (H$-[4.5\mu]$) color map, binned at \SI{5}{\arcsecond} $\times$ \SI{5}{\arcsecond} pixels.
	    \textbf{Middle left:} Mean $\beta$ map.
			The scale goes from 2.01 (black) to 2.06 (white).
			Here we interpret the apparent lack of structure in our map of the mean $\beta$ values as not inconsistent with a lack of spatial variation of the extinction law.
    	\textbf{Middle right:} Mean extinction A(K\textsubscript{S}) map.
			The scale goes from 0.3 (black, low extinction) to 9.5 (white, high extinction).
	    \textbf{Bottom left:} Standard deviation of $\beta$ map.
			The scale goes from 0.28 (black) to 0.295 (white).
			The standard deviation of the $\beta$ values is tight and shows no obvious spatial structure which is not inconsistent with the interpretation of no spatial variation of the extinction law at the GC.
        \textbf{Bottom right:} Standard deviation in A(K\textsubscript{S}) map.
			The scale goes from 0.4 (black) to 0.95 (white).
			Note that the areas of highest extinction (the black cells in \textbf{Middle Left}) have elevated (brighter blue or white) standard deviations, indicating map cells with very few stars in H band or incomplete $[4.5\mu]$ data.
			}% eo caption
        \label{fig:H_H45color}
	    \label{fig:AKsBetaMeans}
\end{figure}

We present our extinction map in Figure~\ref{fig:H_H45color}.
This figure shows the H band image, (H$-[4.5\mu]$) color map, A(K\textsubscript{S}) map, $\beta$ map, and the standard deviation of each pixel's \num{29000} realizations for both A(K\textsubscript{S}) and $\beta$.
Taking note of the bimodality in Figure~\ref{fig:H_H45color}, we examined the histograms of the extinction and $\beta$ values to look for evidence in each of bimodality and variation respectively in Figure~\ref{fig:histogramMeans}.
We find that while there are `typical` and `high` regions of extinction, both are not inconsistent with being described by a single power law.
We have applied the extinction map to the IPSC and present the de-reddened Color-Magnitude Diagrams (CMDs) in Figure~\ref{fig:CMDs}.
We discuss our results for the extinction A(K\textsubscript{S}), $\beta$, relative extinction values and their comparison to previous work in Table~\ref{tab:ExtinctionComparison}, and the de-reddening correction to the IPSC J, H, K\textsubscript{S} photometry in the following subsections.

\subsection{A(K\textsubscript{S}) at the GC}
\label{subsec:AKs}
We find that the distribution of A(K\textsubscript{S}) is strongly double-peaked, as demonstrated by Figure~\ref{fig:H_H45color}. This bimodality is also evident in the color histogram in Figure~\ref{fig:histogramMeans}.
Map cells with values of high extinction, A(K\textsubscript{S})$>7.5$, are presented as a lower limit; these map cells have large H band magnitudes (faint flux) and correspond to the darkest regions in the H band image.
This is reflected in the $\sigma_{K_S}$ map: The regions of highest extinction show the highest $\sigma_{K_S}$ as well.

The standard deviation map of A(K\textsubscript{S}) shows quite a bit of structure.
The typical standard deviation values range in value from \SIrange{0.5}{1.0}{mag}.
The tiger stripes are distinct as are patchy sections to the North and West (upper left), which we interpret as evidence that these regions are so highly extincted that there aren't many stars in those H band map cells.

Given the obvious spatial grouping of high-extinction map cells, we were curious to see if 1) there was a preferred spatial scale in the extinction features, and 2) if there was a difference in the spatial scale distribution of the `typical' and `high' extinction cells.
We ran a 2-D cross-correlation analysis on our extinction map to look for peaks in the distribution of extinction at various spatial scales.
We made binary maps created by setting map cells with A(K\textsubscript{S}$>4$ and $>7$ equal to 1 (and zero everywhere else) to emphasize the `typical' and `high' extinction regimes.
What we found was that there is no preferred angular scale at either typical and high extinction regimes; modulo normalization, there also appears to be no significant difference between the two regimes in the results of the cross-correlation analysis.
We conclude that there is no preferred spatial scale in the extinction features even though they are grouped spatially (e.g., the tiger stripes), and interpret this as being not inconsistent with a uniform extinction law at the GC.

\cite{galacticnucleus2017} uses their holographic imaging data to measure extinction at the GC at a much finer scale than our technique is capable of over a field of view of roughly \si{8}{\arcminute} $\times$ \SI{3.5}{\arcminute} centered on Sgr A*.
However, our results seem to be in good overall agreement, once the difference in platescale is taken into account.
We determined this by accounting for the total extinction value in similar-sized regions in several places where our extinction maps overlap, and found them to be within $1\sigma$ agreement. 

\subsection{$\beta$ at the GC}
\label{subsec:beta}
%fix this line's placement before submission
We investigated whether or not the extremely dark regions, or `tiger stripes' were better described by a different extinction law coefficient, but found no statistically significant difference from the coefficient for regions with less extinction.
In other words, our values for $\beta$ are not inconsistent with a uniform extinction law.
Additionally, there is very little structure evident in the standard deviation map of $\beta$, perhaps surprising given the obvious structure in the standard deviation map of A(K\textsubscript{S}).
We interpret this as evidence supporting our conclusion that our data are not inconsistent with a single power law describing extinction in the GC.
A possible limitation is that the `tiger stripes' and other very H-band dark regions is a lack of H-band flux; there simply aren't many stars in those map cells. This may indicate that the `tiger stripes' are closer to Earth than the GC as we would expect foreground stars to `fill in' the tiger stripes, but this was not pursued farther.

\begin{figure}[hbt]
	\centering
	\begin{minipage}{0.47\textwidth}
		\includegraphics[width=\textwidth]{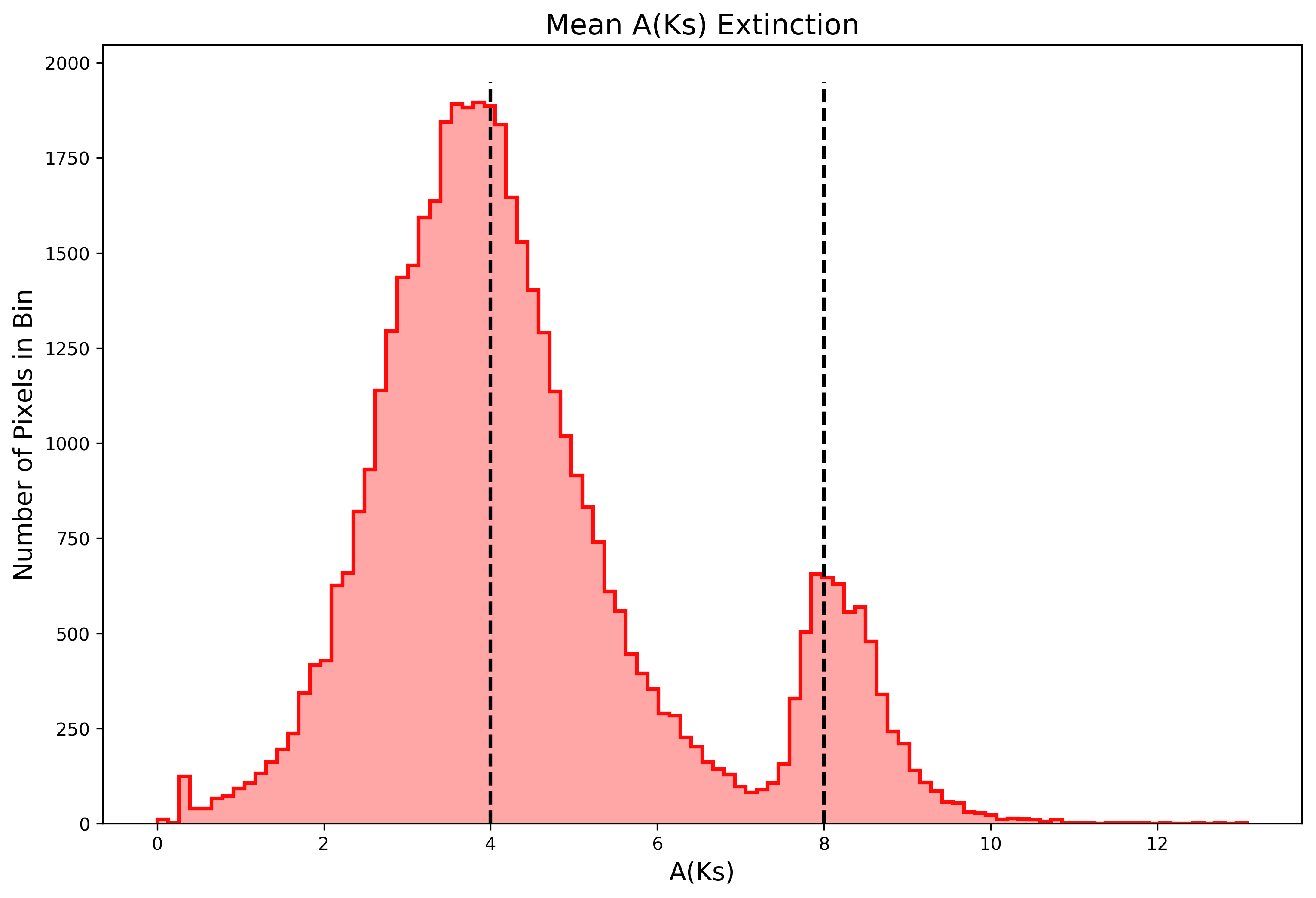}
	\end{minipage}
~
	\begin{minipage}{0.47\textwidth}
		\includegraphics[width=\textwidth]{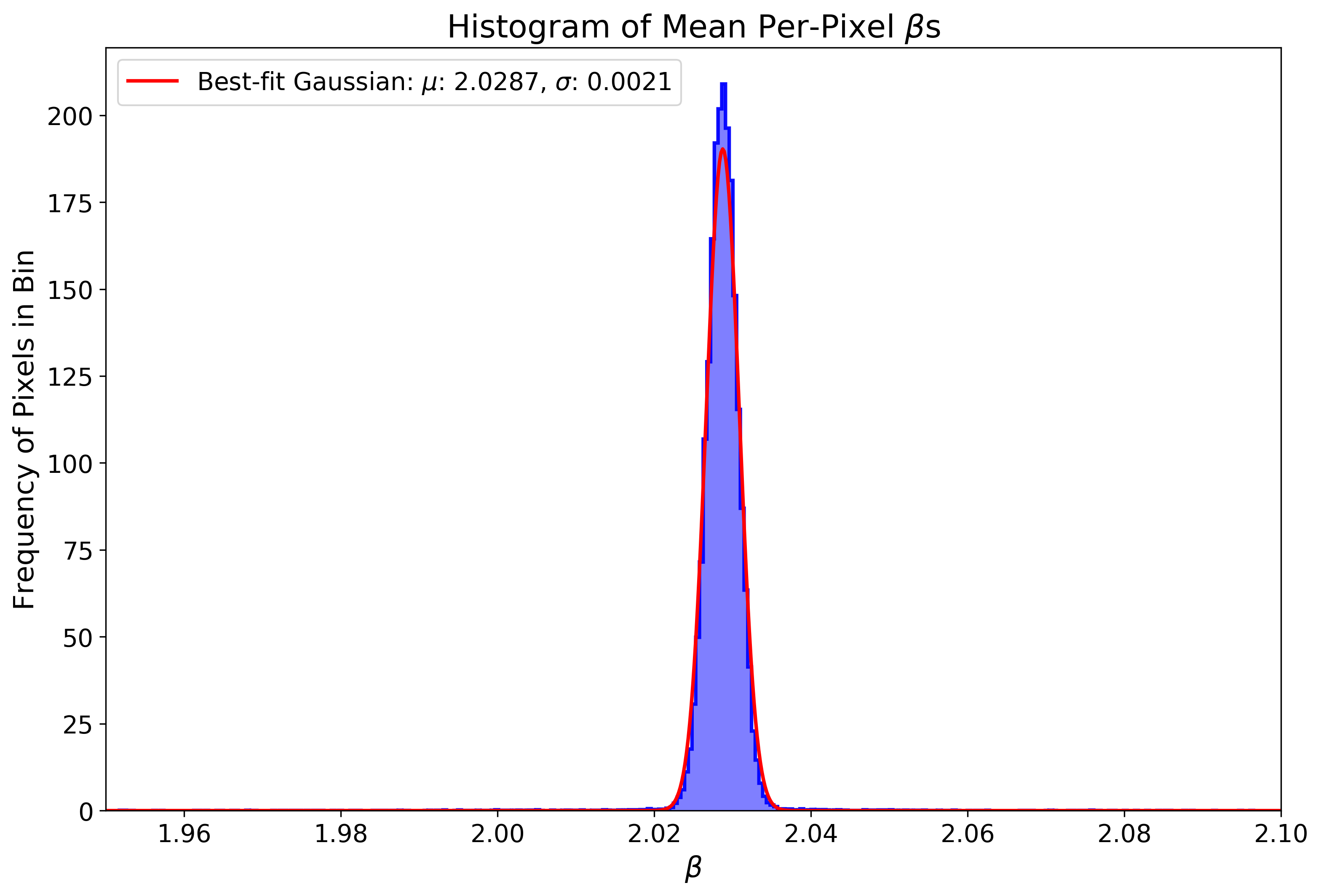}
	\end{minipage}
	\caption{
		Histograms of the mean extinction and $\beta$ values. \textbf{Left:} Histogram of the mean extinction values.
		There is clear evidence of two populations of extinction, one with a mean of about 4 and standard deviation of 1, and the second with a mean of about 8 with a standard deviation of about 0.75; the vertical dotted lines are at A(K\textsubscript{S}) = 4 and 8.
		\textbf{Right:} Histogram of the mean $\beta$ values.
		The peak of the distribution of mean $\beta$s is 2.029 and the standard deviation is 0.002.
		Despite the bimodal distribution of extinction, there is no evidence of a bimodal extinction law. 
	} % eo caption
	\label{fig:histogramMeans}
\end{figure}

We find that the value of $\beta$ does not vary spatially over our entire mapping region of \SI{17}{\arcminute} $\times$ \SI{16}{\arcminute}.
We found the mean value of $\beta$ for each pixel and show the distribution in Figure~\ref{fig:histogramMeans}.
The mean $\beta$ is \num{2.029}, with $\sigma_\beta=0.002$ being the standard deviation of the distribution of mean per-pixel $\beta$ values over the entire GC region we examined.
Individual cells have a rather large $\sigma_\beta$ of about \num{0.3}, but assuming Poisson statistics, we reduce the error on $\beta$ to \num{0.06} per pixel ($0.3/\sqrt{25}=0.06$).
We use \num{25} as the number of map cells in this case because we are only using \num{25} cells in each run; while the global behavior of $\beta$ is self-consistent, we only used \num{25} map cells in our partial pooling model for any given run, and we choose to err on the conservative side when propagating errors.

We conclude that $\beta = 2.029\pm(0.002+0.06)$, where 0.002 is the error of the mean $\beta$ values, and 0.06 the error on individual cells in each set of \num{25}.
Adding the errors in quadrature, we find that $ \beta = 2.029 \pm 0.06$.
This value of $\beta$ is consistent within $1\sigma$ with many of the previous studies using only NIR point-source photometry that found $\beta\simeq2$ in the direction of the GC~\citep{indebetouw2005,Nishiyama2006,Nishiyama2009,Fritz2011}.

We compare our values of $\beta$, $\frac{A(J)}{A(K_s)}$,$\frac{A(H)}{A(K_s)}$, and $C$ to 6 selected works on extinction at the GC in Table~\ref{tab:ExtinctionComparison}.

\cite{indebetouw2005} found that $C=0.61\pm0.04$, $\beta_1 = 2.22\pm0.17$, and $\beta_2 = 1.21\pm0.23$ if one used Equation~\ref{eqn:ExtincSquared}. 
Much of the literature quotes~\cite{indebetouw2005} to have a power law index $\beta=1.65$ or $1.66$~\citep{Nishiyama2006,dustyVeilIII}.
Despite repeated readings of this paper, we were unable to find a single mention of this value or indeed any attempt to fit their data using Equation~\ref{eqn:simpleExtinct}.
However, using the relative extinction values in this paper one can find $\beta$ by fitting their relative extinction values to $A(\lambda)\propto\lambda^{-\beta}$, which is perhaps where $\beta=1.66$ originates; the inconsistency in the cited $\beta$ value in the literature is unclear but might be due to authors' varied preferences in fitting routines.
Our value of $\beta$ is consistent with \cite{Nishiyama2006, Nishiyama2009, Fritz2011} within $<1\sigma$, and is consistent with \cite{Hosek2018} within $1.7\sigma$.

Conversely, \cite{galacticnucleus2017} uses five methods to measure the extinction at the GC with their high-resolution data and find that $\beta=2.31\pm0.03$, which is inconsistent with our measurement at the $3.1\sigma$ level.
However, their quoted extinction value is the average of all of their methods, while the error given is the scatter about the mean of these measurements, which likely underestimates their error (albeit by a small amount). Averaging their measurements and using the errors of those measurements to propagate the error gives $\beta=2.31\pm0.04$, a small difference but one that reduces the discrepancy from $3.1\sigma$ to $2.8\sigma$.
We thus conclude that our measurements are in tension, but not inconsistent, with \cite{galacticnucleus2017}.

\begin{table}[htp]
	\centering
	\caption{NIR extinction law comparison
	}% eo caption
	\label{tab:ExtinctionComparison}
	% !TEX root = ./../../GCExtinction.tex

%% 
%\small{
\setlength{\tabcolsep}{6pt}
\begin{threeparttable}
    \begin{tabular}{ m{1.55in} m{1in} m{1in} m{1in} m{1.in}}
    \toprule
    Publication & $\beta$ & $\frac{A(J)}{A(K_s)}$ & $\frac{A(H)}{A(K_s)}$ & $C$  \\
    \midrule 
    \cite{indebetouw2005}      & $\beta_1=2.22\pm0.17$, $\beta\approx1.66$\tnote{*}      & $2.5\pm0.15$          & $1.55\pm0.08$         & $0.61\pm0.04$                       \\
    \cite{Nishiyama2006}       & $1.99\pm0.02$         & $3.021\pm0.004$       & $1.73\pm0.01$         & $0.494\pm0.006$                     \\
    \cite{Nishiyama2009}       & $2.23\pm0.23$\tnote{\dag}         & $3.02\pm0.04$         & $1.73\pm0.03$         & \ldots\tnote{\ddag}                               \\
    \cite{Fritz2011}           & $2.11\pm0.06$         & \ldots                & $1.74\pm0.14$\tnote{\S}         & \ldots                              \\
    \cite{galacticnucleus2017} & $2.31\pm0.03$\tnote{\P}         & \ldots                & \ldots                & \ldots                              \\
    \cite{Hosek2018}           & $2.38\pm0.15$\tnote{\maltese}     & $3.69$      & $1.99$       & \ldots \\
    This work                  & $2.03\pm0.06$         & $2.57\pm0.03$         & $1.42\pm0.04$         & $0.60$\tnote{\copyright}                     \\
    \bottomrule
    \end{tabular}
    \begin{tablenotes}
        \item [*] \cite{indebetouw2005} uses a polynomial extinction law as in Equation~\ref{eqn:ExtincSquared}, but one can fit a simple power law to their data and get $\beta\approx1.66$ as is often quoted in the literature.
        \item [\dag] We use the value of $\beta$ as given in Fritz et al. (2011).
    	\item [\ddag] We use `\ldots' to denote where cited papers do not explicitly give the values used in this table and are not straightforwardly derived.
    	\item [\S] We derived the $\frac{A(H)}{A(K_s)}$ for Fritz et al. (2011) from their given $A(H)$ and $A(K_s)$ values.
    	\item [\P] See \ref{subsec:beta} for a discussion of \cite{galacticnucleus2017}'s $\beta$ value to ours.
    	\item [\maltese] Hosek et al. (2018) note that a power-law fit does not describe their NIR extinction measurements toward Westerlund 1 and RC stars at the GC, and instead opt for a $3^{rd}$ degree spline fit from \SIrange{0.8}{2.14}{\micro\meter}.
    	\item [\copyright] We fixed the intercept value to be $0.60$.
    \end{tablenotes}
\end{threeparttable}

\end{table}

The fact that the mean $\beta$ values across our map are so tightly distributed is strong evidence that the extinction law toward the GC is invariant spatially despite the strong double-peaked distribution of the extinction.
The lack of spatial variance of the extinction law at the GC is also consistent with previous work~\citep{Nishiyama2006,Nishiyama2009,Foster2013,galacticnucleus2017}.

\cite{Hosek2018} stresses in their discussion that their NIR extinction curve is best fit with a $3^{rd}$ degree spline fit in place of a power law, but if one insists on fitting their data to a power law, $\beta=2.38\pm0.15$, a $1.7\sigma$ difference from our value of $\beta$.

\subsection{Relative Extinction at the GC}
\label{subsec:relExtinct}
Given the extinction law described by Equation~\ref{eqn:simpleExtinct} and our value of $\beta$, we present the relative extinctions in the NIR and comparisons to two other studies in Table~\ref{tab:ExtinctionComparison}.
To find the errors in our relative extinction values we differentiated Equation~\ref{eqn:simpleExtinct} with respect to $\beta$ and used \num{0.06} as our error in $\beta$.
%\vspace{0.5ex}

%\end{landscape}

The relative extinction values calculated by~\cite{indebetouw2005} are within $0.5\sigma$ and $1.7\sigma$ for A(H/A(K\textsubscript{S}) and A(J)/A(K\textsubscript{S}), respectively, of our results despite \cite{indebetouw2005} using the RC method for the relative extinction values and using a different IR extinction law (they favored using a second-order polynomial, as in Equation~\ref{eqn:ExtincSquared}). 
Their power-law index, $\beta_1$, is consistent with our $\beta$ value at the $1\sigma$ level.
Comparing our results to~\cite{Nishiyama2006} shows an interesting mix of agreement and disagreement.
Our $\beta$ values are consistent at the $1\sigma$ level, but our relative extinction and power law intercepts are substantially inconsistent.
The difference is in large part due to the different intercept values we use for our respective power laws, and our methods for measuring the extinction are also quite different; they performed point-source photometry on stars in the GC which naturally restricted them to stars that are most likely on the near side of the GC, whereas we used surface brightness photometry and took advantage of the severe crowding at the GC.
Effectively, the 'last photosphere of extinction' that we measure is closer to the GC than the stars Nishiyama et al. were able to perform point source photometry on in J, H, and K\textsubscript{S}.

\cite{Fritz2011} gives A(H) and A(K\textsubscript{S}), so we include their value in Table~\ref{tab:ExtinctionComparison}; we are consistent within $1.8\sigma$.
Other previous works do not explicitly give relative extinctions.

\subsection{De-Reddened IPSC Color-Magnitude and Color-Color Diagrams for J, H, K\textsubscript{S} Photometry}
\begin{figure}[hbt]
	\centering
	\begin{minipage}{0.42\textwidth}
		\includegraphics[width=\textwidth]{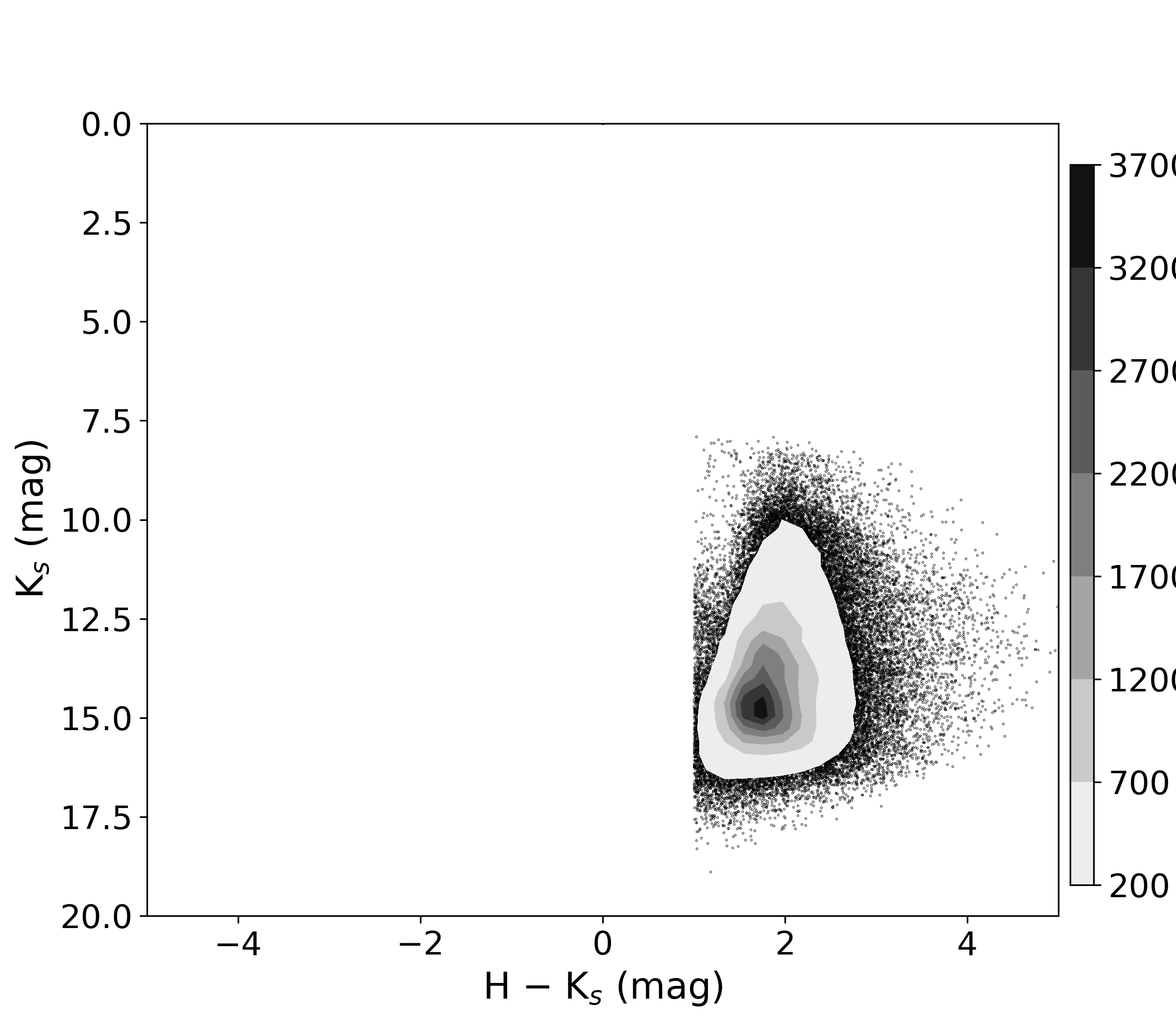}
	\end{minipage}
~~
	\begin{minipage}{0.42\textwidth}
		\includegraphics[width=\textwidth]{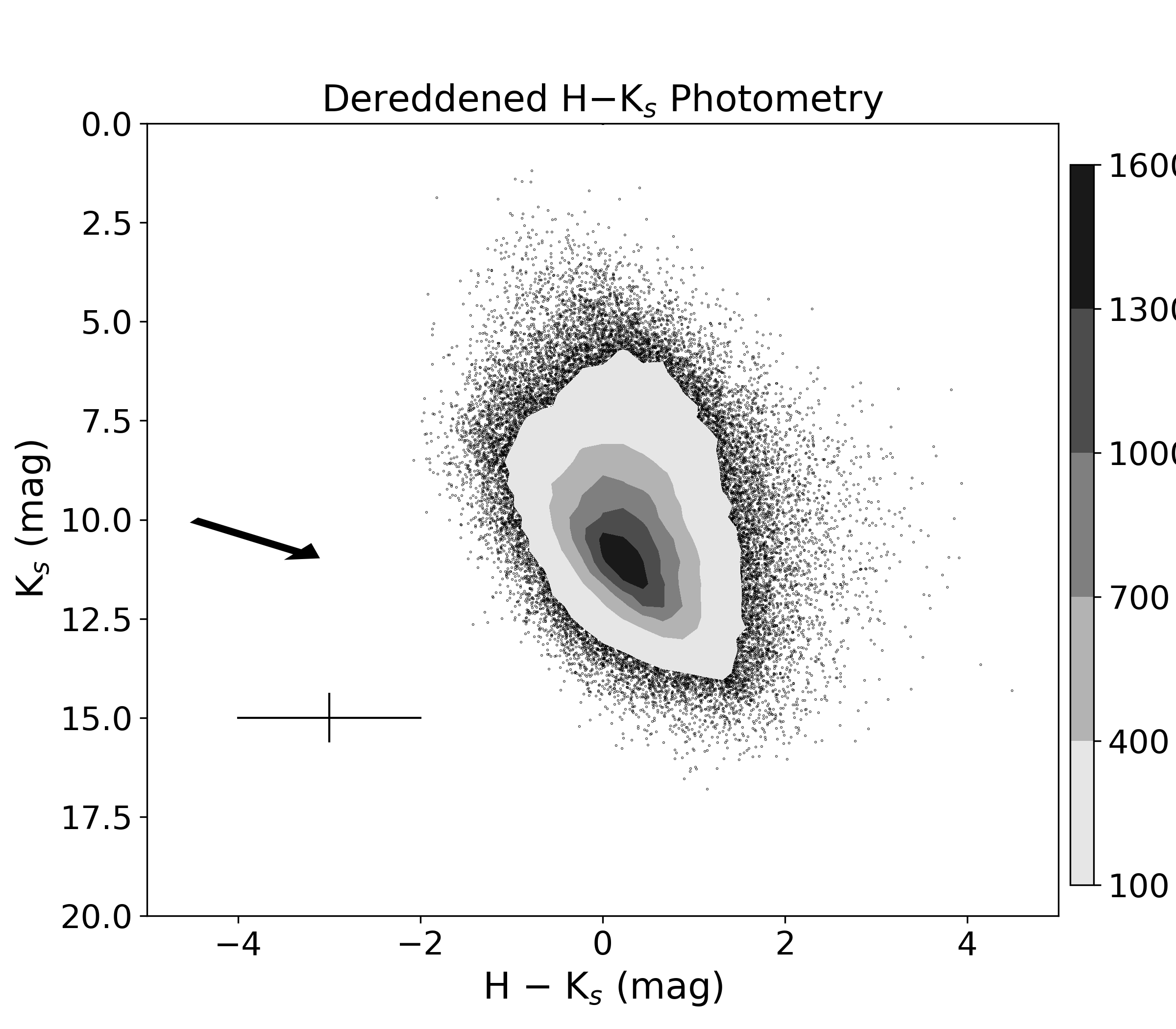}
	\end{minipage}
\\	
	\begin{minipage}{0.42\textwidth}
		\includegraphics[width=\textwidth]{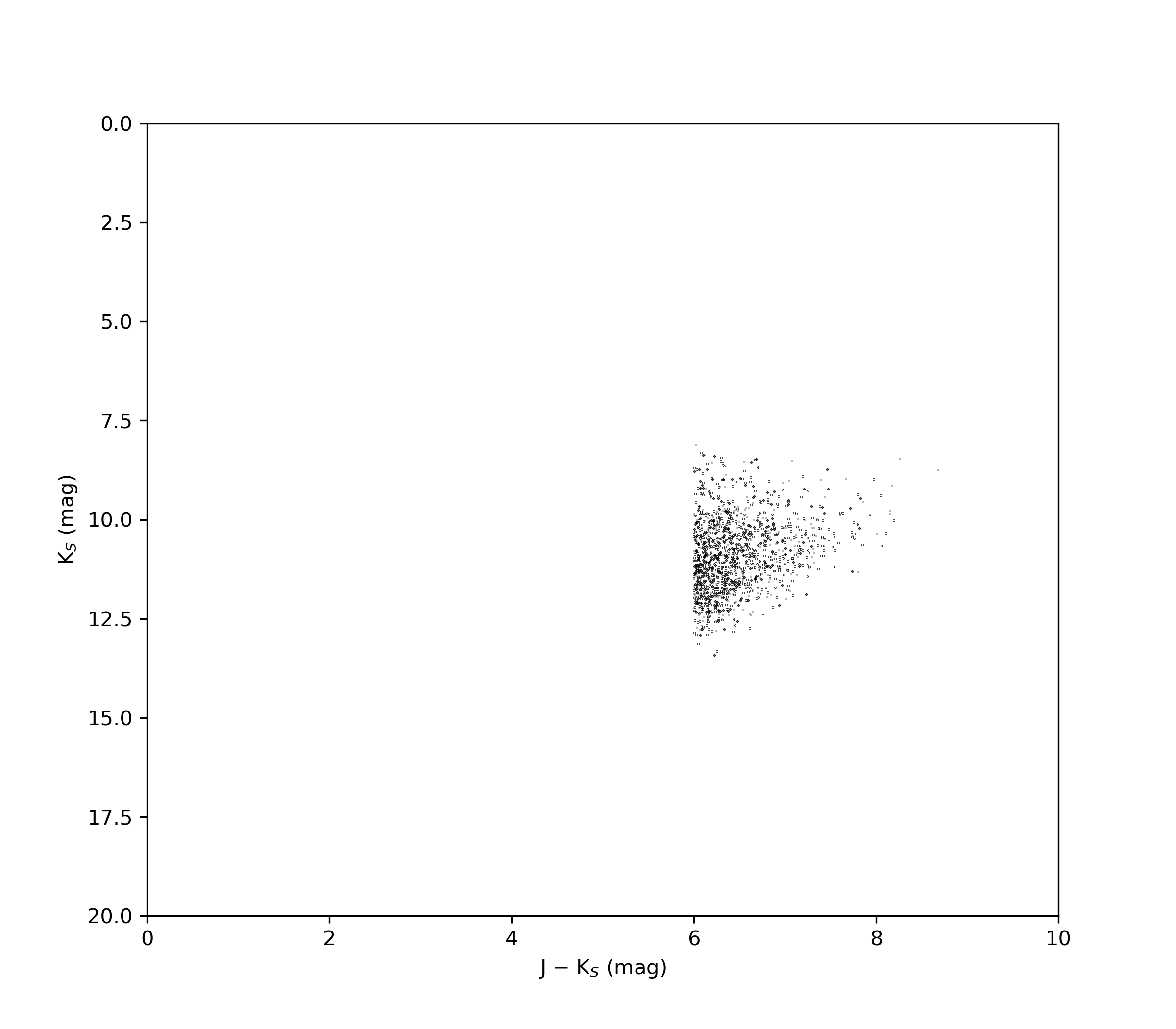}
	\end{minipage}
~~
    \begin{minipage}{0.42\textwidth}
		\includegraphics[width=\textwidth]{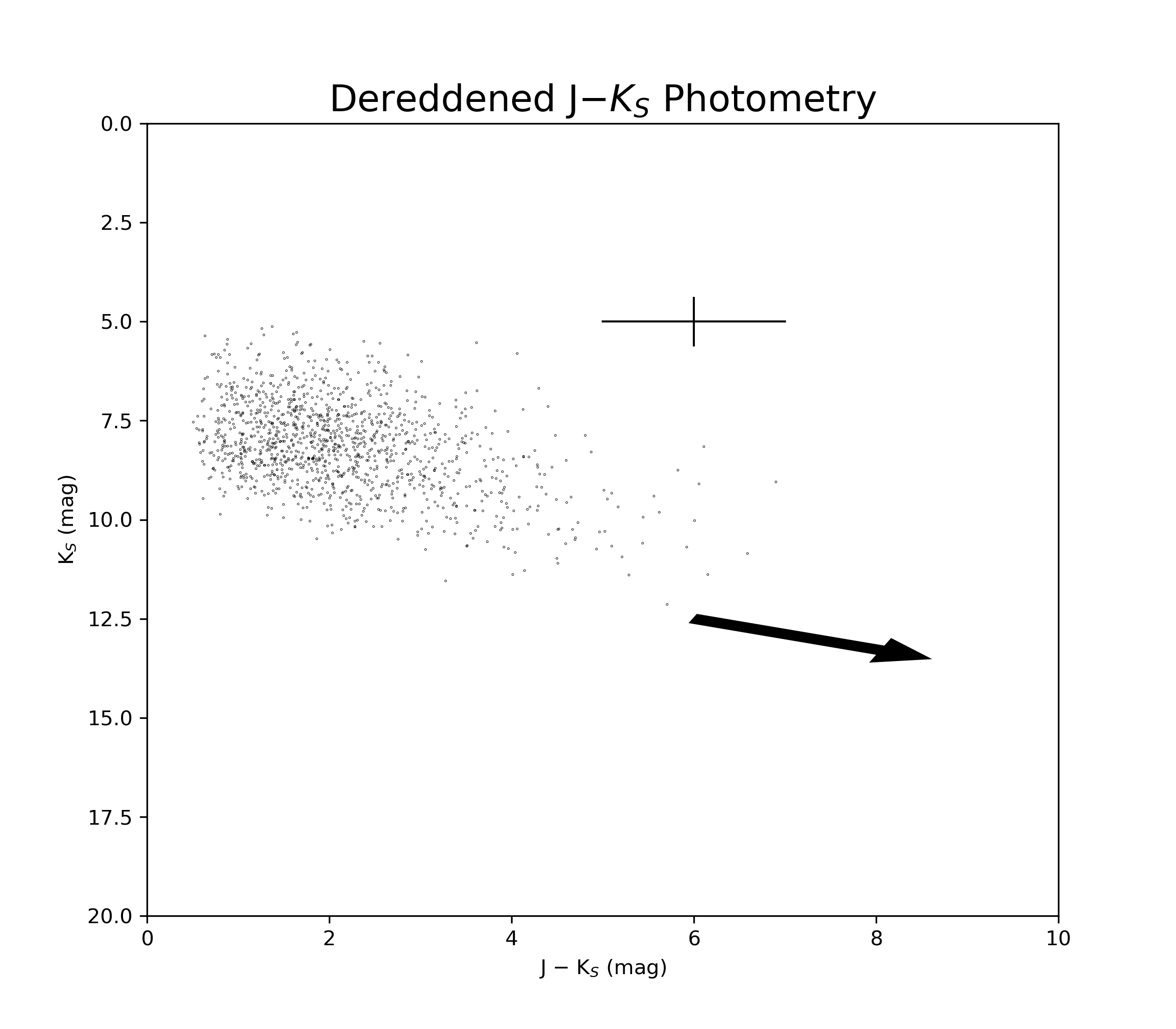}
	\end{minipage}
\\
	\begin{minipage}{0.42\textwidth}
		\includegraphics[width=\textwidth]{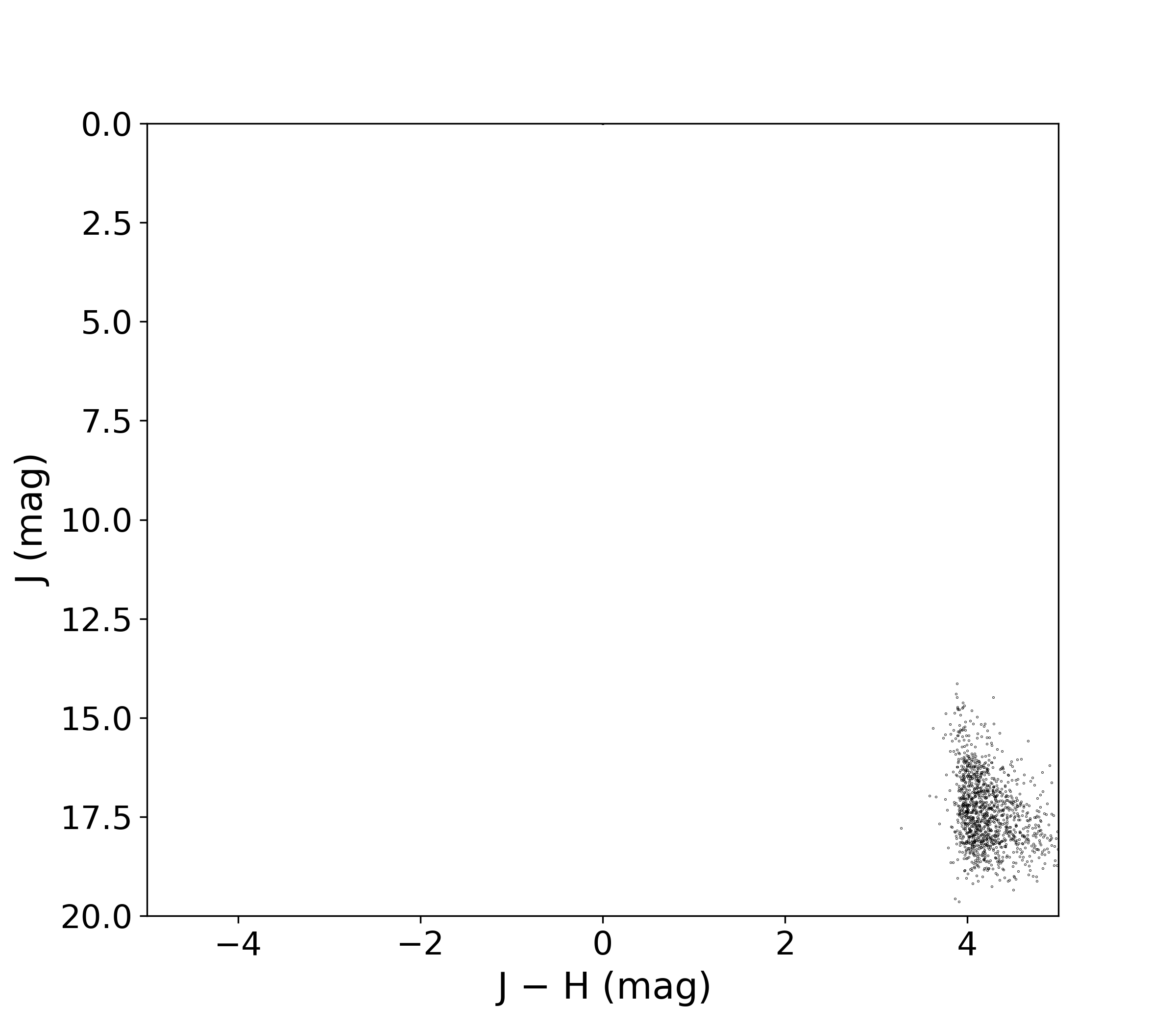}
	\end{minipage}
~~
    \begin{minipage}{0.42\textwidth}
		\includegraphics[width=\textwidth]{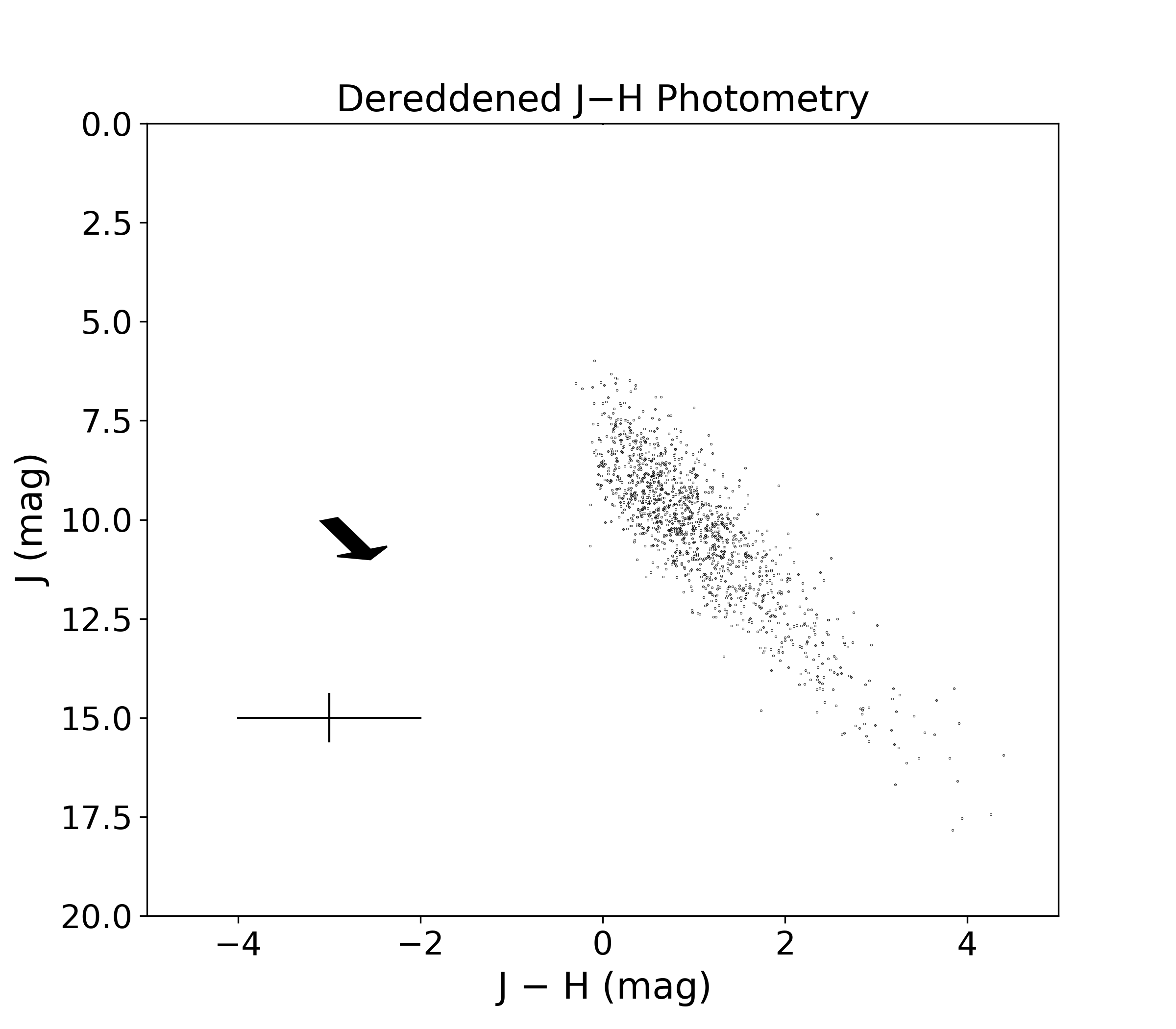}
	\end{minipage}
	\caption{
	    The left column plots are un-dereddened CMD diagrams while the right column plots are de-reddened CMD diagrams using our extinction map.
	    Each de-reddened diagram has representative error bars derived from Figure~\ref{fig:dereddeningsigmas} in the ordinate and abscissa, respectively.
	    The errors from the IPSC are orders of magnitude smaller and are not shown.
        \textbf{Top:} H $-$K\textsubscript{S} CMD for \num{124089} stars.
	    The black arrows show the reddening vector arising from \SI{1}{mag} of extinction in K\textsubscript{S}.
        \textbf{Middle:} K\textsubscript{S} vs J - K\textsubscript{S} CMD for \num{13178} stars.
        The reddening vector shown is \SI{1}{mag} of extinction in K\textsubscript{S}.
        \textbf{Bottom:} J vs J - H CMD for \num{13178} stars.
        The reddening vector is shown as a black arrow and is \SI{1}{mag} of extinction in J (equivalent to \SI{0.39}{mag} of extinction in K\textsubscript{S})
	}% eo caption
		\label{fig:CMDs}
\end{figure}

\begin{figure}[htpb]
    \centering
	\begin{minipage}{0.47\textwidth}
		\includegraphics[width=\textwidth]{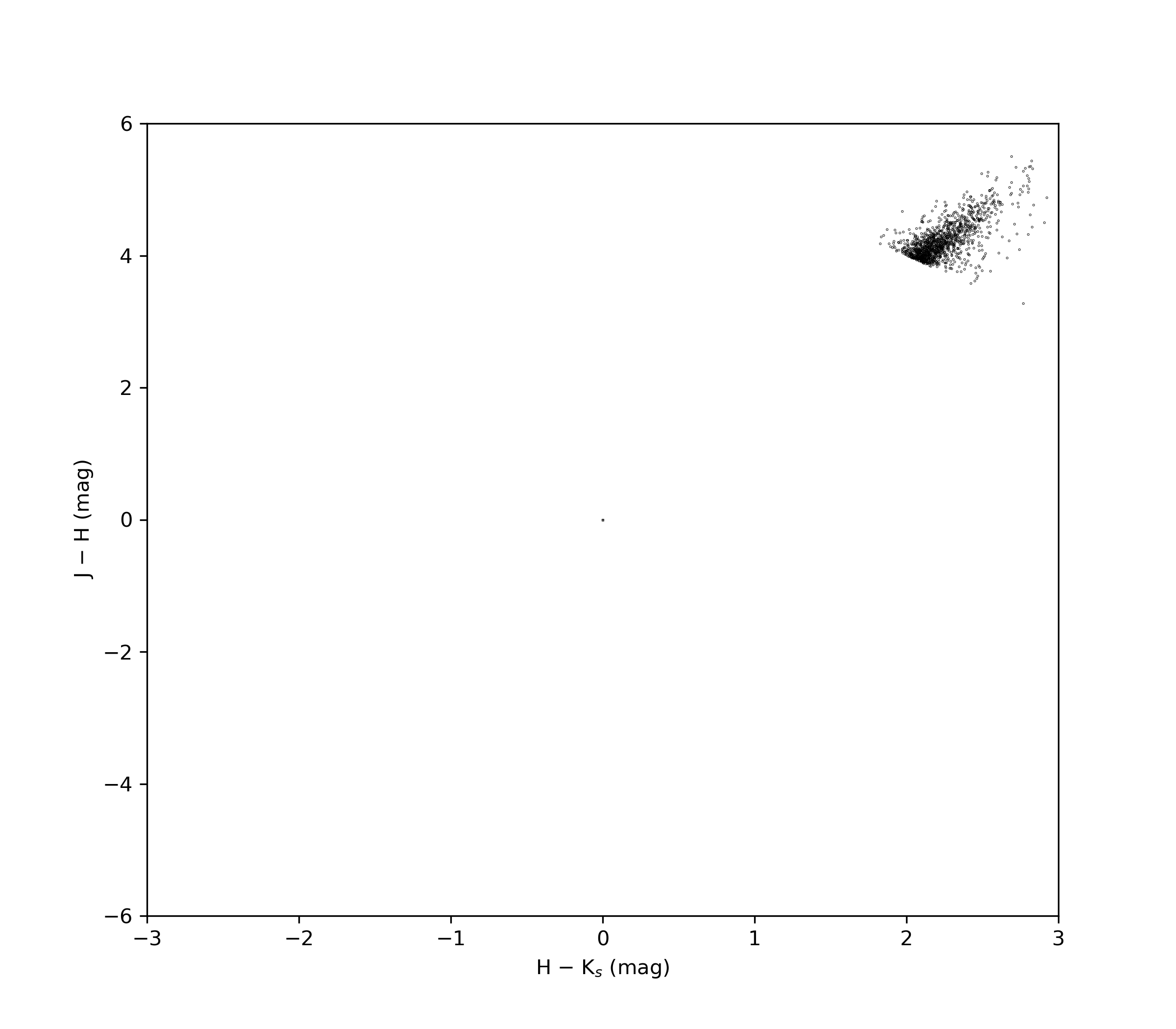}
	\end{minipage} 
~~
\begin{minipage}{0.47\textwidth}
		\includegraphics[width=\textwidth]{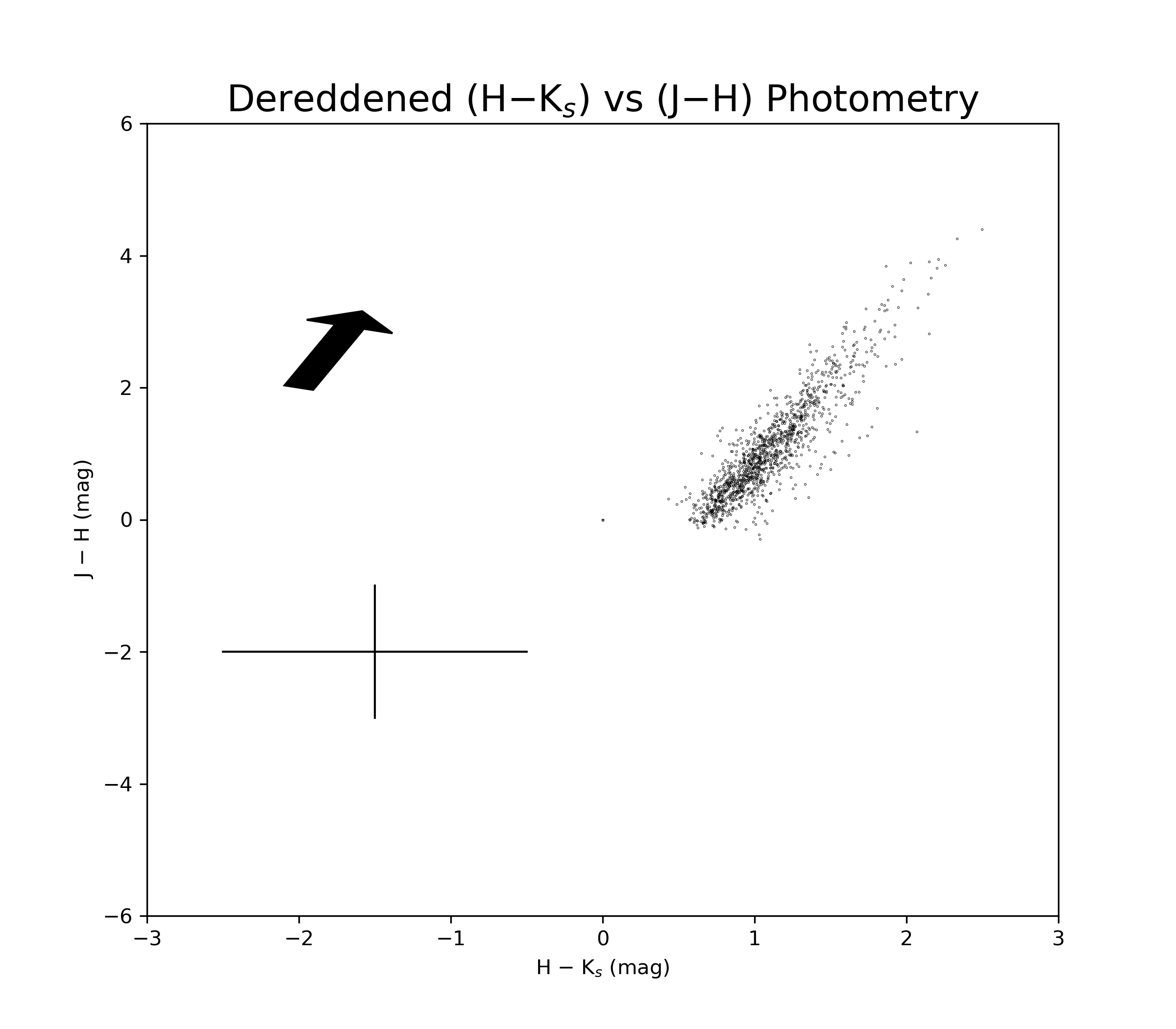}
	\end{minipage}
	\caption{Color-color diagrams of (J - H) vs (H - K\textsubscript{S}) using \num{13178} stars. 
	  The error bar shows the typical dereddening error of $\pm$\SI{1}{mag} in K\textsubscript{S} and $\pm$\SI{1.0}{mag} in (J$-$H) and (J$-$K\textsubscript{S}) derived from Figure~\ref{fig:dereddeningsigmas}.
	  \textbf{Left:} Un-dereddened (J - H) vs (H - K\textsubscript{S}) photometry. 
	  \textbf{Right:} Dereddened (J - H) vs (H - K\textsubscript{S}) photometry. 
	  The reddening vector is shown as a black arrow and is \SI{1}{mag} of extinction in K\textsubscript{S}.
	} % eo caption
	\label{fig:colorcolor}
\end{figure}

In order to test our extinction map, we applied the extinction map to the IPSC and present the un-deredded and de-reddened Color-Mag\-ni\-tude Diagrams (CMDs) in Figure~\ref{fig:CMDs} and the un-dereddened and de-redded color-color diagram in Figure~\ref{fig:colorcolor}.
In order to construct the CMDs, we performed two sets of color and extinction cuts to the IPSC using our extinction map and value of $\beta$ to calculate the amount of de-reddening to apply to stars in each pixel map.
To make the (H$-$ K\textsubscript{S}) vs  K\textsubscript{S} CMD, we required stars to have both H and K\textsubscript{S} photometry, leaving us with \num{137451} stars.
We then made a moderate color cut, keeping only stars with (H$-$K\textsubscript{S})$>1.0$; stars with a bluer (smaller) color are most likely foreground stars as even intrinsically blue stars at the GC are reddened by several magnitudes in K\textsubscript{S}, and have an (H$-$K\textsubscript{S}) color $>1$ as a result.

To find the appropriate color cut to use in constructing our (H$-$K\textsubscript{S}) CMD, we considered a star at the GC behind A(K\textsubscript{S})$=$5 mag of extinction with an intrinsic (H$-$K\textsubscript{S})\textsubscript{0} color of \SI{-1}{mag}; its observed (H$-$K\textsubscript{S}) color is about \SI{1}{mag} because the color correction term (A(H) $-$ A(K\textsubscript{S})) is about 2, using our relative extinction value A(H)/A(K\textsubscript{S})$=$\SI{1.42}{mag}.
This implies that even intrinsically blue stars at the GC are not removed from consideration with a color cut of (H$-$K\textsubscript{S})$>1$.
We also ignore stars which are located in map cells with A(K\textsubscript{S})$>7.5$ because these stars are most likely in front of IRDCs and so bias our measurement, over-correcting the reddening of these individual stars.
Figure~\ref{fig:CMDs} shows the un-dereddened CMD and the dereddened CMD using the color and magnitude cuts. 
We constructed a color-color diagram and show it in Figure~\ref{fig:colorcolor}.

\begin{figure}[htb]
	\centering
    \begin{minipage}{0.47\textwidth}
        \includegraphics[width=\textwidth]{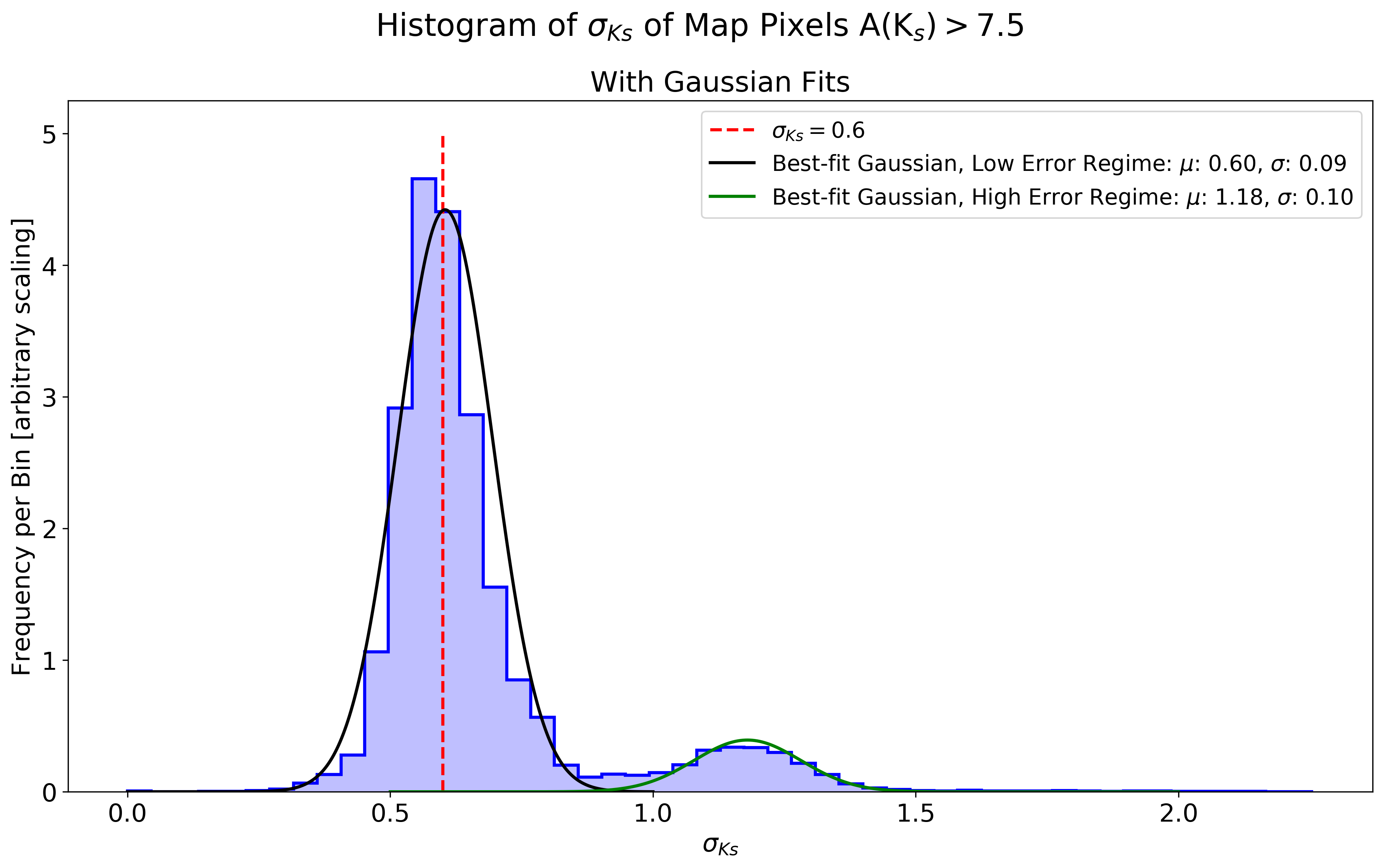}
    \end{minipage}	
~~
	\begin{minipage}{0.47\textwidth}
	    \includegraphics[width=\textwidth]{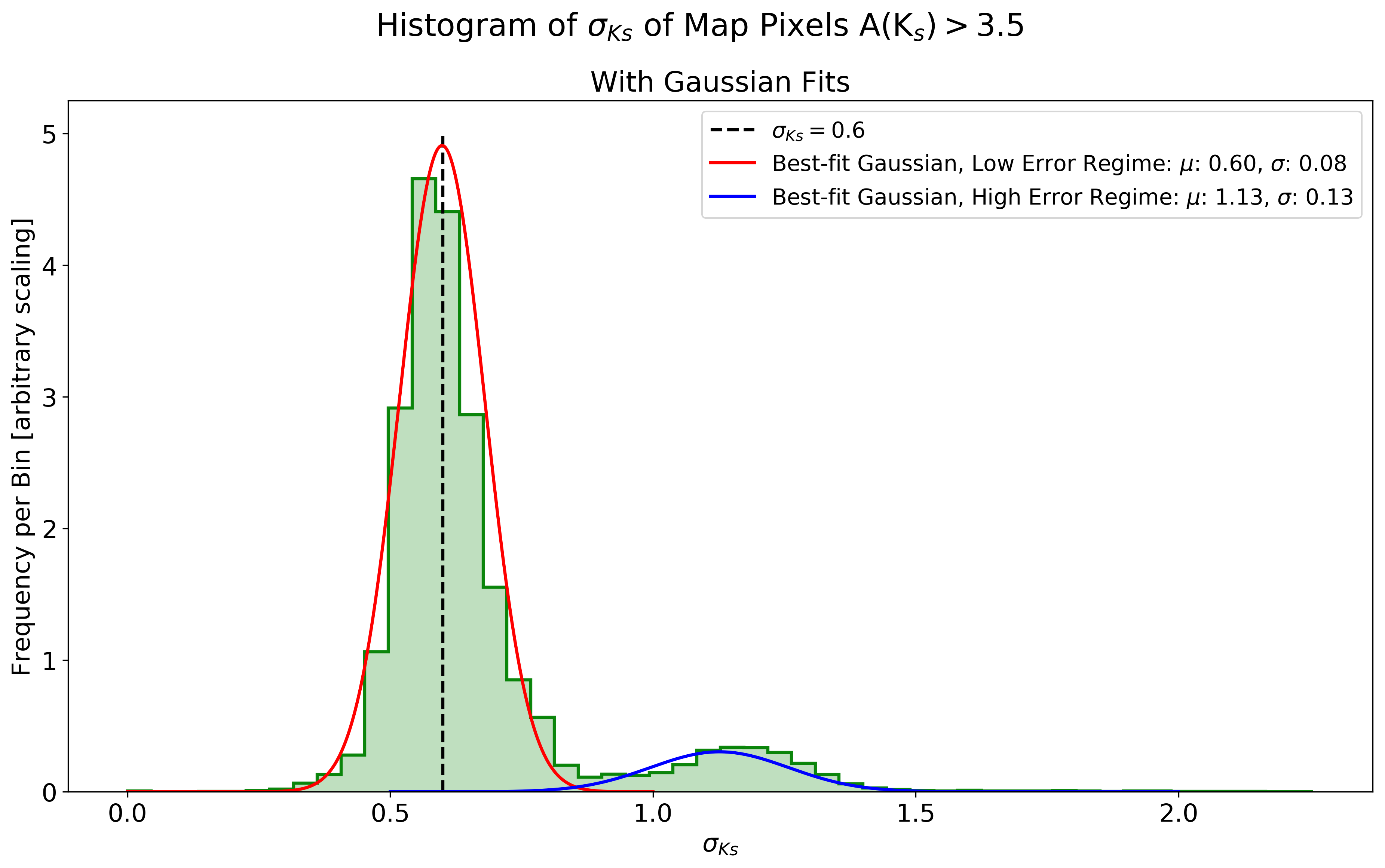}
    \end{minipage}
    \caption{
        Histogram of map pixel $\sigma_{A(Ks)}$ for H$-$K\textsubscript{S}, J$-$K\textsubscript{S}, and J$-$H de-reddening.
	    Each star is assigned its error based on the standard deviation of A(K\textsubscript{S}) over the     \num{29000} traces.
	    \textbf{Left:} We had \num{124089} stars after our magnitude and color cuts.
		The typical error is about \SI{0.6}{mag}, but for some map cells the error is as high as \SI{1.5}{mag}.
		\textbf{Right:} Histogram of map pixel $\sigma_{A(Ks)}$ for  dereddening for the \num{13178} stars that make the cut for J band.
		The typical error is about \SI{0.6}{mag}, but for some map cells the error is as high as \SI{1.25}{mag}.
	} %eo caption
	\label{fig:dereddeningsigmas}
\end{figure}

We do not show the errors of our dereddening in the CMD plots to reduce clutter, but do have representative error bars shown in the lower left (the photometric errors from the IPSC are orders of magnitude smaller and neglected as a result).
In order to determine the errors in H and K\textsubscript{S}, we examined the histogram of the errors in A(K\textsubscript{S}) in Figure~\ref{fig:dereddeningsigmas}, which includes stars from the IPSC that have both H and K\textsubscript{S} magnitudes.
We note that each star is assigned its error based on the standard deviation of A(K\textsubscript{S}) for the \num{29000} realizations for each map pixel.
Based on this histogram, we assign each star an error in A(K\textsubscript{S}) of \SI{0.6}{mag} and an (H$-$K\textsubscript{S}) of \SI{1}{mag}.
We examined a histogram of the errors in A(K\textsubscript{S}) in Figure~\ref{fig:dereddeningsigmas} which included only stars from the IPSC that have measurements in all three photometric bands to determine the error in J.
Based on this histogram, we assign each star an error in (J$-$H) and (J$-$K\textsubscript{S}) of \SI{1}{mag}.

After the color and magnitude cuts, we are left with \num{124089} stars.

Similarly, to make the (J$-$K\textsubscript{S}) vs  K\textsubscript{S} CMD, as well as the color-color diagram (J$-$K\textsubscript{S}) vs (H$-$ K\textsubscript{S}), we selected stars with J, H, and K\textsubscript{S} photometry from the IPSC, giving us \num{32920} point sources (stars).
We next considered a star at the GC located in a map pixel with A(K\textsubscript{S}) of 7.5 with an intrinsic (J$-$K\textsubscript{S})\textsubscript{0} of 0, or about as blue as stars can be.
Using our relative extinction value A(J)/A(K\textsubscript{S})$=$2.57, A(J) is then 19.3 mag, giving this (imaginary) intrinsically blue star a color of (J$-$K\textsubscript{S})$=$11.8.
If we reduced the A(K\textsubscript{S}) threshold to 4 mag, we found that the color correction term (A(J)$-$A(K\textsubscript{S}) is 6.3.
We used a color cut of (J$-$K\textsubscript{S})$>6.0$, and used a magnitude cut of A(K\textsubscript{S})$<3.5$, which left us with \num{13178} point sources for our (J$-$H) vs J and (J$-$K\textsubscript{S}) vs J CMDs, as well as our color-color diagram.

\section{Conclusions and Future Work}
\label{sec:GCFutureWork}

We have presented an extinction map of the inner \SI{16.5}{\arcminute} by \SI{17}{\arcminute} region of the Galaxy using H and $[4.5\mu]$ surface brightness photometry.
We have found the infrared extinction law at the GC to be not inconsistent with being spatially invariant despite the obvious bimodal distribution of extinction values, and that the power law index $\beta = 2.03\pm0.06$.
Using our extinction law map and the ISPI Point Source Catalog, we have also constructed dereddened CMDs, although the shallow (in terms of distance from the Solar System) J band photometry strongly reduces the usefulness the  of the dereddened CMDs made with J band photometry.

\cite{dustyVeilI} note that other NIR-MIR color combinations are worth consideration for extinction mapping, such as (K\textsubscript{S}$-[3.6\mu]$).
As noted previously, broad emission features commonly attributed to PAH emission occurs at MIR wavelengths near \num{3.3}, \num{6.2}, \num{7.7}, and \SI{8.6}{\micro\meter}, limiting the usefulness of MIR bands other than $[4.5\mu]$ in determining extinction.
Given the width of the \emph{Spitzer} IRAC filters, this means that the only filter not subject to PAH contamination is $[4.5\mu]$.
The other filters present an opportunity to test PAH emission as a tracer for extinction by correlating PAH maps derived from other MIR bands to the extinction map.

%: Acknowledgements
\acknowledgments
We gratefully acknowledge Dan Gettings for sharing his \texttt{map\_tools} code, for teaching us how to use it, and many fruitful conversations as to the nature of measurements and modeling in a Bayesian framework.
Our reviewer went above and beyond in their insightful and helpful comments, and this work was made stronger and more approachable because of their effort; all mistakes, from grammatical to technical, remain the fault of the authors.

This research made use of Astropy,\footnote{http://www.astropy.org} a community-developed core Python package for Astronomy \citep{astropy:2013, astropy:2018}.
This work is based in part on observations made with the Spitzer Space Telescope, which is operated by the Jet Propulsion Laboratory, California Institute of Technology under a contract with NASA.

%: Software acknowledgements
\software{TRILEGAL \citep{trilegal2012}; DAOPHOT \citep{stetson1987}; superFATBOY \citep{prelim-sFB2012}; Source Extractor \citep{sextractor1996}; SCAMP \citep{scamp2006}; SWARP \citep{swarp2002}; \texttt{pymc3}  (Salvatier J, Wiecki T.V., Fonnesbeck C., 2016, Probabilistic programming in Python using PyMC3. PeerJ Computer Science 2:e55 \href{https://doi.org/10.7717/peerj-cs.55}{https://doi.org/10.7717/peerj-cs.55}); astropy \citep{astropy:2013,astropy:2018}}

\facility{CTIO, Spitzer (IRAC)}.

\appendix
\restartappendixnumbering
\section{Choosing Priors and \texttt{pymc3}}
\label{sec:GCPriors}
We chose to use a Bayesian approach to our modeling of extinction in order to take advantage of the large amount of measured color excess data.
We used the Python package \texttt{pymc3}\footnote{Salvatier J, Wiecki T.V., Fonnesbeck C. (2016) Probabilistic programming in Python using PyMC3. PeerJ Computer Science 2:e55 \href{https://doi.org/10.7717/peerj-cs.55}{https://doi.org/10.7717/peerj-cs.55}}.
\texttt{pymc3} is a Bayesian statistical modeling suite that uses a flexible Markov chain Monte Carlo approach to modeling.
A discussion of the machinery of \texttt{pymc3} and Bayesian statistics is beyond the scope of this paper, but we will discuss how we constructed our Bayesian model of extinction in this section and note that \texttt{pymc3} uses a highly adaptive, gradient-based sampler, which frees us of having to implement own sampler code.

The basic idea of our model is simple:
Find A(K\textsubscript{S}) given the observed color excess (H$-[4.5\mu]$) and expected (H$-[4.5\mu]$)\textsubscript{0}, and find the range of $\beta$ values that are consistent with the data.

The Bayesian part comes in because we have an intrinsic color spread ($\sigma_{\textrm{H}-[4.5\mu]}$), an expected distribution of $\beta$ values, and a limit on what the extinction can be (e.g., negative A(K\textsubscript{S}) values are not physical, although non-thermal emission from PAH or dust is a possible source of bluer-than-expected color).
We can also leverage partial pooling, which uses a common distribution of errors or estimated values in comparing multiple data points, to better estimate the extinction law.
What this means in our context is that we drew test values for our variables, like $\beta$, from a single distribution for each realization.
A mild complication is that sharp increases in extinction can be due to InfraRed Dark Clouds (IRDCs,~\cite{IRDC2006}), which at the distance of the GC (\SI{8.5}{\kilo\parsec}) are typically about \SI{10}{\arcsecond}, or two map cells, in size and can have extinctions as high as A(K\textsubscript{S}) of \numrange{10}{15}. 
This means that, depending on the extinction law used, IRDCs can have absolute visible extinction(A\textsubscript{V}) values in excess of 100 magnitudes.
The `tiger stripes', filamentary structures apparent to the right of the center of the H-band image (Figure~\ref{fig:hBand}) are very dark, indicating extremely high extinction typical of IRDCs.

\vspace{1ex}
\begin{table}[htp]
	\centering
	% !TEX root = ./../../Thesis.tex

%% 
\begin{threeparttable}[c]
	\caption{
	  Model input parameters for \texttt{pymc3}.} %eo caption
    \begin{tabular}{cccccc}
        \toprule
        	Param                  & Type\tnote{\dag}       & Mean   & $\sigma$ & Lower  & Upper  \\
        \midrule
        	A(K$_{\textrm{S}}$)	   & Uniform\tnote{\ddag}    & \ldots & \ldots   & 0.01   & 15     \\
        	$\sigma_{A(K_{s})}$    & HalfNormal\tnote{\S} & 0      & 0.5      & \ldots & \ldots \\
        	$\beta$                & Uniform    & \ldots & \ldots   & 1.5    & 2.5   \\
        \bottomrule
    \end{tabular}
        \begin{tablenotes}
            \item [\dag] The `Type' column signifies what sort of pre-defined class of variable for parameter definitions in \texttt{pymc3}.
        	\item [\ddag] In the case of a Uniform distribution (which is continuous), we give lower and upper bounds.
        	\item [\S] The HalfNormal class is a continuous normal distribution with a mean of zero and restricted to only positive values; these properties make the HalfNormal useful as an error term.
    \end{tablenotes}
\end{threeparttable}

    \label{tab:ExtinctionParams}
\end{table}

We designed our model with the parameters listed in Table~\ref{tab:ExtinctionParams}.
The model itself is simple. For each pixel we draw an A(K\textsubscript{S}), $\beta$, and $\sigma_{K_S}$ from their respective distributions.
We then calculate $\gamma$ as defined in Equation~\ref{eqn:dustyVeilExtinction}: $1/\left(10^{0.6-\beta\log_{10}\left[\textrm{H}\right]}-10^{0.6-\beta\log_{10}\left[4.5\mu\right]}\right)$, and use that value to calculate the model color excess $M$ based on the extinction and $\gamma$ values:
\begin{math}
M(\textrm{H}-[4.5\mu]) = \gamma*A(K_S) + 0.08
\end{math}

We then compare $M$, the model color excess, to the measured $E(\textrm{H}-[4.5\mu])$ color, using $M$ as the mean and $\sigma_{A(K_s)}$ as the standard deviation of a normal distribution (e.g., we assume Gaussian errors to our measured values).
The model is then fed into \texttt{pymc3}, which we initialized with the No U-Turn Sampler (or NUTS), an auto-adaptive Hamiltonian sampler written expressly for \texttt{pymc3}.
We set the sampler to run \num{30000} realizations, and initialized it with \num{100000} `samples' that use auto-differential variational inference (ADVI) to find and define the gradient for the parameter space of the model; this both finds initial values and sets the step size for each parameter appropriately.
We recorded the traces using the HDF5 file format and analyzed them with the python package \texttt{pytables} and our custom-written code.

\subsection{Splitting up the Data}
A significant constraint for \texttt{pymc3} and other partial pooling MCMC code is that it runs progressively slower with larger datasets.
We found that running our model with the full dataset was impossible (each realization in our Monte Carlo chain took approximately one CPU-day).
We explored two ways of splitting up the data into subsets; while this diminishes the power of Bayesian partial pooling, we were able to complete \num{30000} realizations for each realization in less than 2 CPU-weeks as opposed to \num{30000} CPU-days.

\begin{figure}[htb]
	\centering
	\begin{minipage}{0.48\textwidth}
		\includegraphics[width=\textwidth]{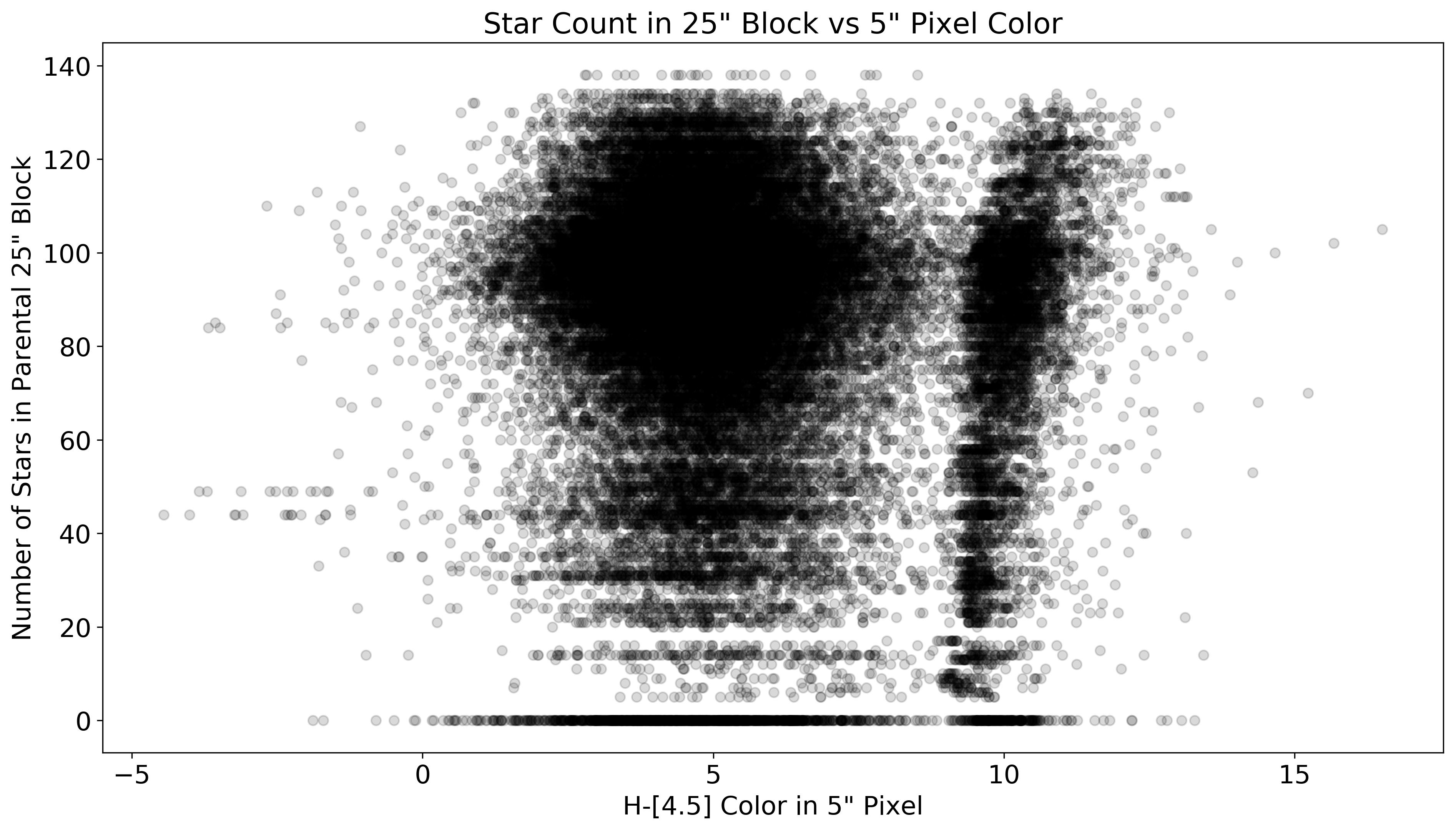}
	\end{minipage}
	\hfill
	\begin{minipage}{0.48\textwidth}
		\includegraphics[width=\textwidth]{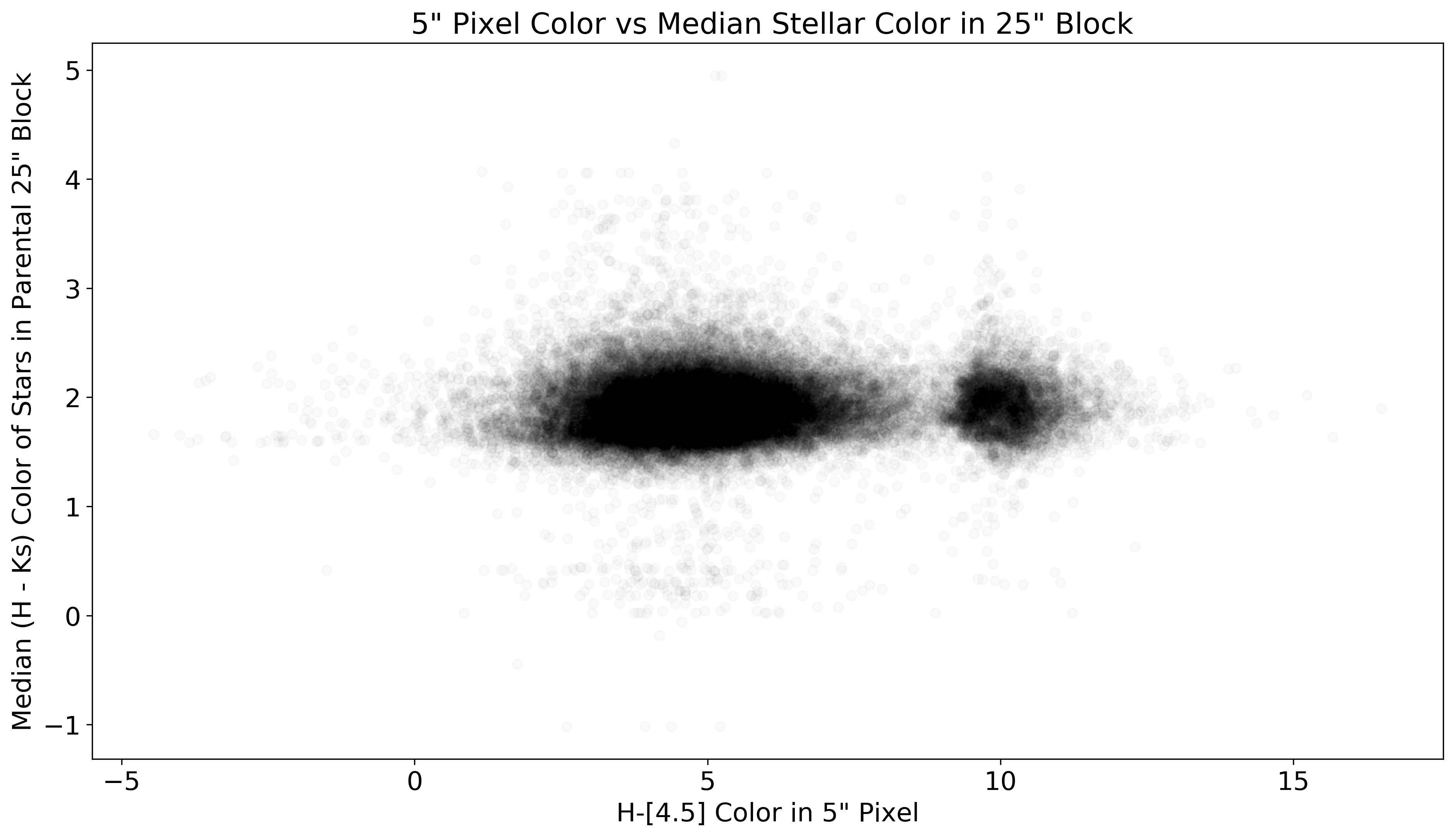}
	\end{minipage}
	\vspace{0.5ex}
	\caption{
		Clustering of map cells.
		\textbf{Left:} Number of stars detected in K\textsubscript{S} in the IPSC within each \SI{25}{\arcsecond}x\SI{25}{\arcsecond} parental box versus the (H$-[4.5\mu]$) color of each pixel in that parental box.
		Note that there are two seemingly distinct clusters of points separated at about (H$-[4.5\mu]$)=8.5.
		\textbf{Right:}  Same as above, but with the median (H-K\textsubscript{S}) color of stars within each parental box (the photometry taken from the IPSC) versus the (H$-[4.5\mu]$) color of each pixel in that parental box.
	} % eo caption
	\label{fig:25ascluster}
\end{figure}

The first way we split the data was spatially; we grouped map cells into subsets of 100, gridding up our map into \num{10} cell $\times$ \num{10} cell `parental boxes.'
While computationally efficient and conceptually straightforward, we found that the standard deviations of the extinction values of these parental boxes showed stark boundary differences at the borders; neighboring parental boxes had notably different means and standard deviations.
This clearly indicated that we were enforcing artificial structure in our calculations of the extinction at the GC.

We looked more closely at our data, looking specifically at each of the \num{10} cell $\times$ \num{10} cell parental boxes and the star counts from the IPSC within each one.
When we plotted the number of stars present in the IPSC in K\textsubscript{S} in the parental boxes versus the color of each pixel (assigning the number of stars in the larger \num{10} $\times$ \num{10} `parental' boxes to each pixel in that box), we found that the data appeared to cluster nicely into two regimes by eye.
Smaller parental boxes of \num{5} $\times$ \num{5} map cells showed similar clustering (as shown in Figure~\ref{fig:25ascluster}, top panel) and when we looked at the median (H$-$K\textsubscript{S}) of all stars from the IPSC in the parental boxes versus the color of each pixel, we found similar clustering (Figure~\ref{fig:25ascluster}, bottom panel).

There are three clusters apparent in Figure~\ref{fig:25ascluster}:
\begin{enumerate}
    \item There are a few map cells that contain \num{0} stars in the IPSC.
    \item Most map cells have stars and have an H$-[4.5\mu]$ color $<8.5$.
    \item The remainder cells have a color H$-[4.5\mu]>8.5$.
\end{enumerate}

As such, we elected to cut the data into two populations split at (H$-[4.5\mu]$) color = 8.5, with the \textbf{A} population having an (H$-[4.5\mu]$) color $\leq 8.5$ and the \textbf{B} population having an (H$-[4.5\mu]$) color $>8.5$.

\subsection{Turning the Crank}
After dividing up the data into two populations, we then randomly selected 25 map cells of either \textbf{A} or \textbf{B} populations and ran the \texttt{pymc3} code, recording each realization in HDF5 files.
Each run (consisting of 25 randomly-selected cells from solely either the \textbf{A} or \textbf{B} populations) took approximately 15 minutes using one CPU core.
We then recombined the data by reading in each HDF5 file, recording the mean and standard deviation of the last \num{29000} realizations for A(K\textsubscript{S}) and $\beta$; the first \num{1000} realizations we discarded to ensure that we used realizations that were no longer using the initial values for our variables, a practice known as `burn-in.'
The total computation time for this analysis was slightly over 2 CPU-weeks.

We performed two sets of cross-checks to verify that we were using both enough map cells and using enough realizations in our MCMC chains.
We ran the data independently of each other for our checks with 100 total cells each: one set used a sample of 100 cells and the other used four samples of 25 cells each.
The four samples of 25 cells were drawn randomly from the sample of 100 cells in order to isolate any possible selection effects from the cross-checks we performed.
The differences were taken on a per-pixel basis, which is why we chose to draw the 4 smaller samples of 25 cells each from the larger sample of 100 cells.
\begin{figure}[htb]
	\begin{center}
	\begin{minipage}{0.44\textwidth}
		\includegraphics[width=\textwidth]{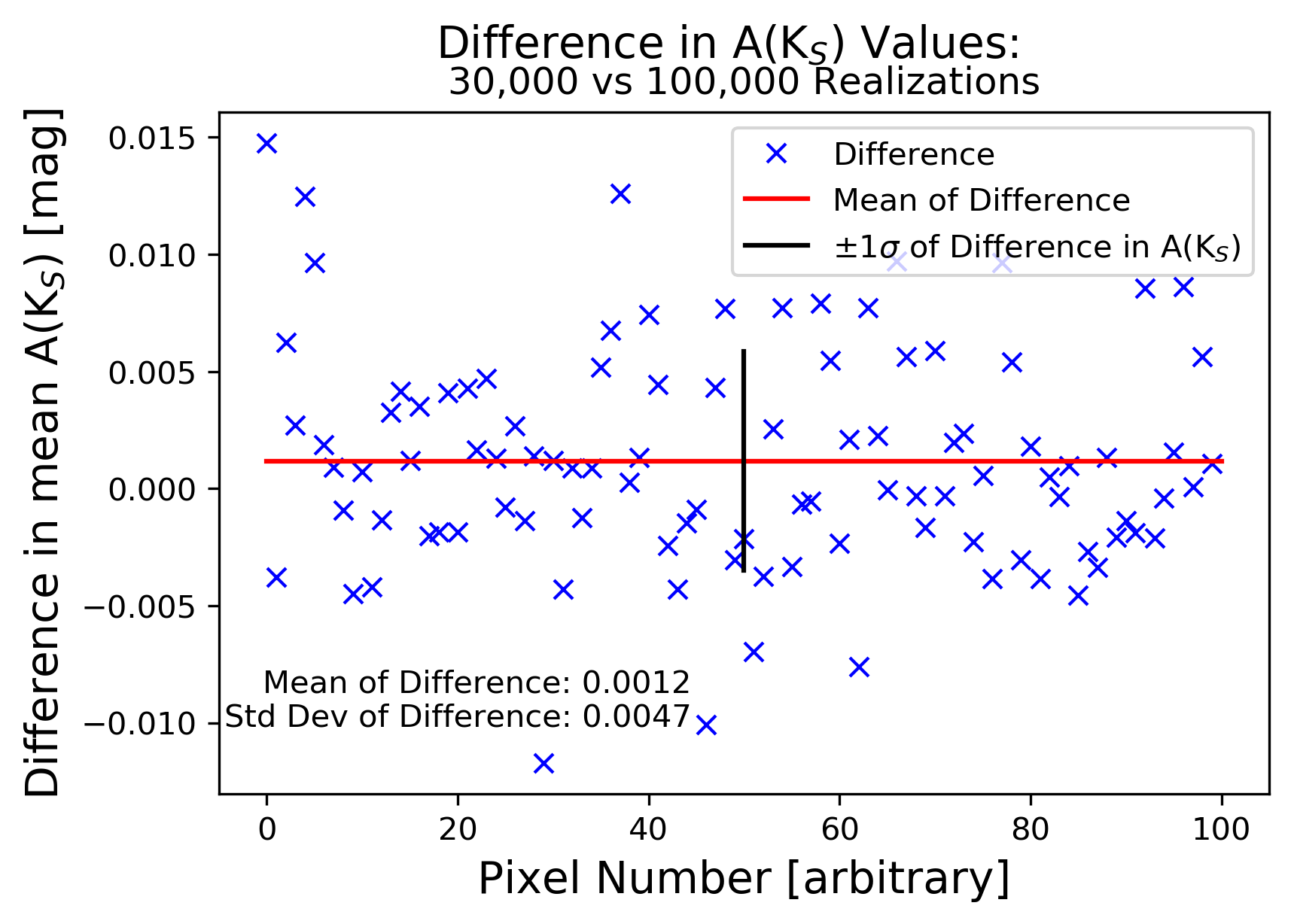}
	\end{minipage}
	~
	\begin{minipage}{0.44\textwidth}
		\includegraphics[width=\textwidth]{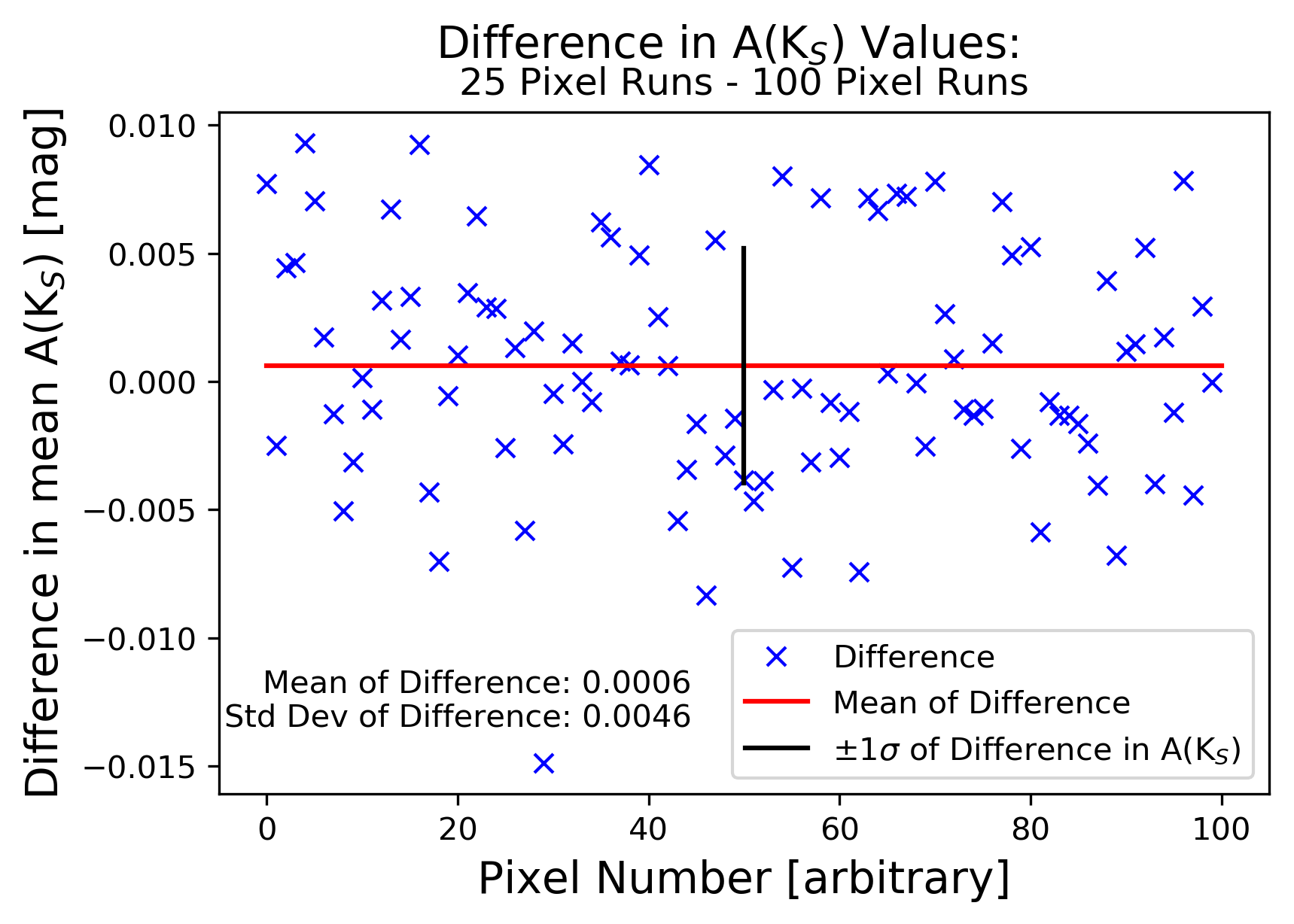}
	\end{minipage}
	\end{center}
	\caption{
	  \texttt{pymc3} cross-checks.
	  Note the Y axis range covers only \SI{25}{millimag}.
	  \textbf{Left:} The mean A(K\textsubscript{S}) values for 4 sets of 25 randomly-chosen pixels using \num{30000} runs minus the mean A(K\textsubscript{S}) values for the same 100 pixels as 1 set using \num{100000} runs.
	  The red horizontal line shows the average difference while the black vertical line represents $\pm1$ standard deviation of the difference in the means. 
	  \textbf{Right:}  As above, but this time comparing the mean  A(K\textsubscript{S}) values for 4 sets of 25 randomly-chosen pixels using \num{30000} realizations minus the mean A(K\textsubscript{S}) values for the same \num{100} pixels as 1 set using \num{30000} realizations.
  	}% eo caption
    \label{fig:pymc3Comparison}
\end{figure}

\subsubsection{Sensitivity to number of realizations}
To check if we had a good convergence of our MCMC chains, we ran chains of \num{100000} realizations for both the four sets of 25 map cells each and the set of 100 cells.
We compared the results of the four smaller samples of 25 cells against the larger sample of 100 cells.
We found that some variation of the mean A(K\textsubscript{S}) did occur at the $\sim0.05\sigma_{K_S}$ level, which is not unexpected given the quasi-random sampling method of \texttt{pymc3}'s NUTS sampler.
Similar variations occurred  for $\beta$.
Figure~\ref{fig:pymc3Comparison} shows two panels; the first panel shows the difference in A(K\textsubscript{S}) between the group of \num{100} pixels vs the four groups of \num{25} pixels.
There are slight differences between the sets of realizations; however, the variation of both A(K\textsubscript{S}) and $\beta$ is small and consistent with 0.
As a result, we conclude that using \num{30000} realizations in our MCMC modeling is sufficient for our purposes.

\subsubsection{Sensitivity to the sample size}
As a test to see if we were using too small a sample of pixels per draw (25), we ran tests with a sample of \num{100} pixels for \num{30000} realizations and compared it to four groups of \num{25} pixels each.

We found that the sample of 100 pixels did have slight but statistically insignificant difference in the means of A(K\textsubscript{S}) on a per-pixel basis, as shown in Figure~\ref{fig:pymc3Comparison}.
The results for $\beta$ were similarly insensitive to sample size.

\bibliography{GCbib}

\end{document}